\documentclass[11pt]{article}

\RequirePackage{latexsym}
\RequirePackage{graphicx}
\RequirePackage{amssymb}
\RequirePackage{natbib}
\RequirePackage{setspace}

\usepackage{color}
\usepackage{array}
\usepackage{hyphenat}

\begin{document}
\definecolor{gold}{rgb}{0.85,0.66,0}
\definecolor{grey}{rgb}{0.5,0.5,0.5}
\definecolor{blue}{rgb}{0,0,0}
\definecolor{red}{rgb}{0,0,0}

     \begin{figure}
     \begin{center}
     \includegraphics[scale=1]{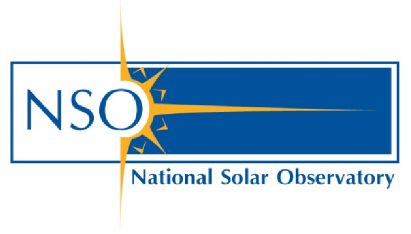}
     \end{center}
     \end{figure}
\title{{\bf Zeemanfit:  Use and Development of the solis\_vsm\_zeemanfit code}
\\$\,$}
\author{Anna L.~H.~Hughes, Jack Harvey, Andrew R. Marble, Alexei A.~Pevtsov, 
\\ \& the SOLIS Pipeline Working Group of the NSO Integrated Synoptic Program
\\$\,$
\\ Institution: National Solar Observatory Integrated Synoptic Program
\\$\,$  \\$\,$  \\$\,$}
\maketitle

\noindent\rule{\linewidth}{0.2mm}
\rule{\linewidth}{0.5mm}

\begin{center}
Technical Report No. {\bf NSO/NISP-2013-02}
\\$\,$
\end{center}

\begin{abstract}

The purpose of the Zeemanfit Code is to provide a straight-forward, easily checked measure of the total magnetic-field strength in the high-strength umbral regions of the solar disk.  In the highest-strength regions, the Zeeman splitting of the 6302\AA\, Fe line becomes wide enough for the triplet nature of the line to be visible by eye in non-polarized light.  Therefore, a three-line fit to the spectra should, in principle, provide a fairly robust measure of the total magnetic-field strength.

The code uses the Level-1.5 spec-cube data of the SOLIS-VSM 6302-vector observations (specifically the Stokes-I and Stokes-V components) to fit the line profiles at each appropriate pixel and calculate the magnetic-field-strength from the line-center separation of the two fit 6302.5 $\sigma$-components.  The 6301.5\AA \ Fe line is also present and fit in the VSM 6302-vector data, but it is an anomalous-Zeeman line with a weaker response to magnetic fields.  Therefore, no magnetic-field measure is derived from this portion of the spectral fit.

\end{abstract}

\pagebreak
\tableofcontents

\section[\textcolor{blue}{Base Data and Product}]{\textcolor{blue}{Base Data and Product}}
\label{Data}

\begin{figure}
     \begin{center}
     \includegraphics[scale=0.22]{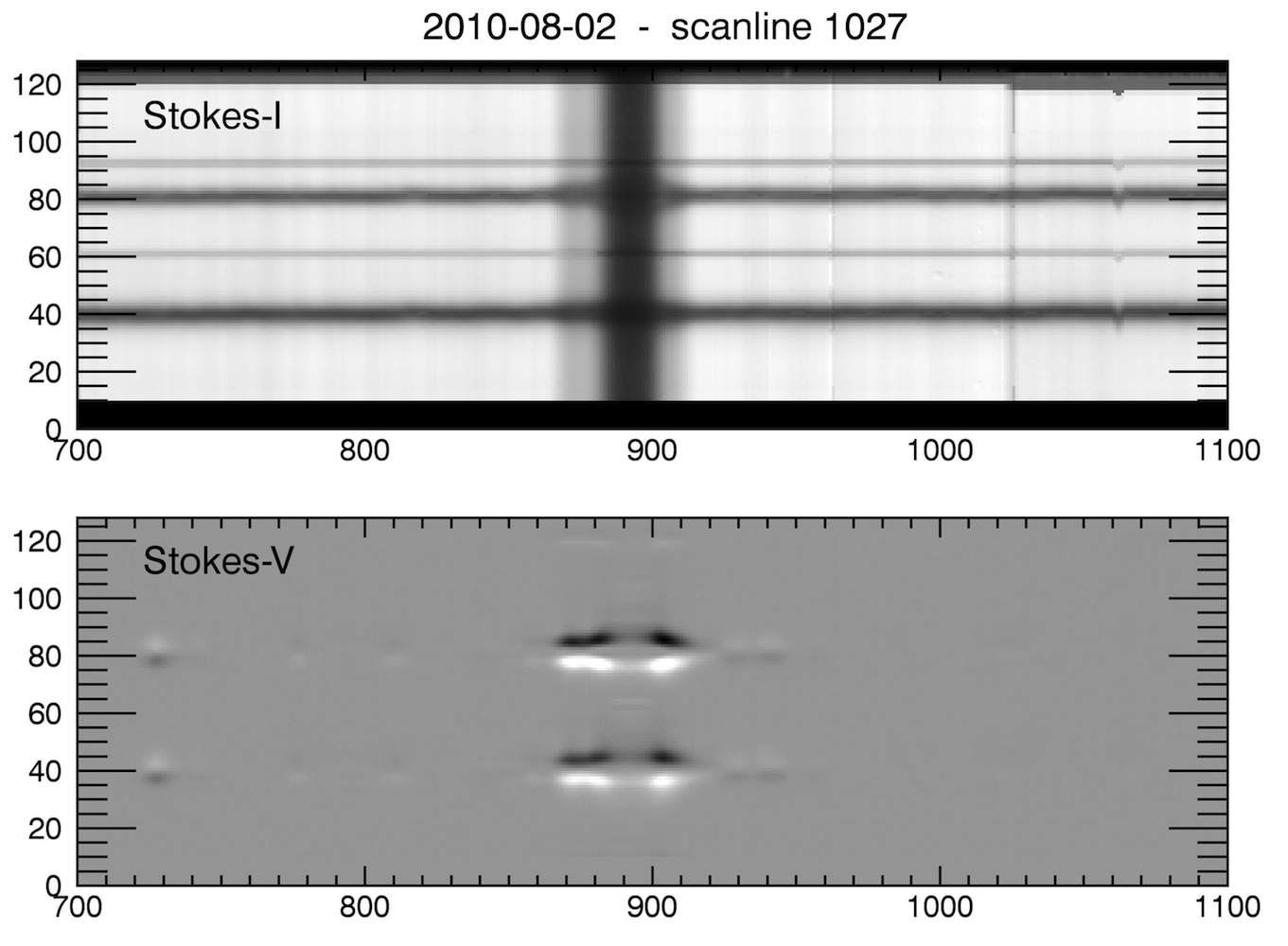}
     \caption[Example Stokes-I and Stokes-V scan-line data]{Examples of Level-0.5 scan-line data running across an active region.  The images' horizontal axes correspond to pixel columns of the disk-image pixels.  The vertical axes denote pixels along the spectral-data plane.}
     \label{FIG_05data}
     \end{center}
     \end{figure}

The base data ingested by the Zeemanfit code consists of the cleaned, Level-0.5 or Level 1.5 VSM 6302v (and 6302l for PROVER2 = 1.2) spectral data (Stokes-I and Stokes-V components only), such as that shown in Figure~\ref{FIG_05data}.
While the code will function using only Level-0.5 scanline data, optimally it takes in Level-1.5 distortion-corrected spec-cube data, along with Level-2-processed continuum data, which is used to aid in sorting pixels to be Zeeman-fit (\S\ref{Code_Sorting}) and to provide initial guesses for the Stokes-I line-center-positions to the MPFITFUN() fitting routine (\S\ref{Code_TripletPos}).  By using Level-1.5/2-processed data, the code is able to produce the most accurate results for a more complete set of active-region pixels across the solar disk and to output those results in the clearest format, including a frame of the continuum image as a reference.

\begin{figure}
     \begin{center}
     \includegraphics[scale=0.2]{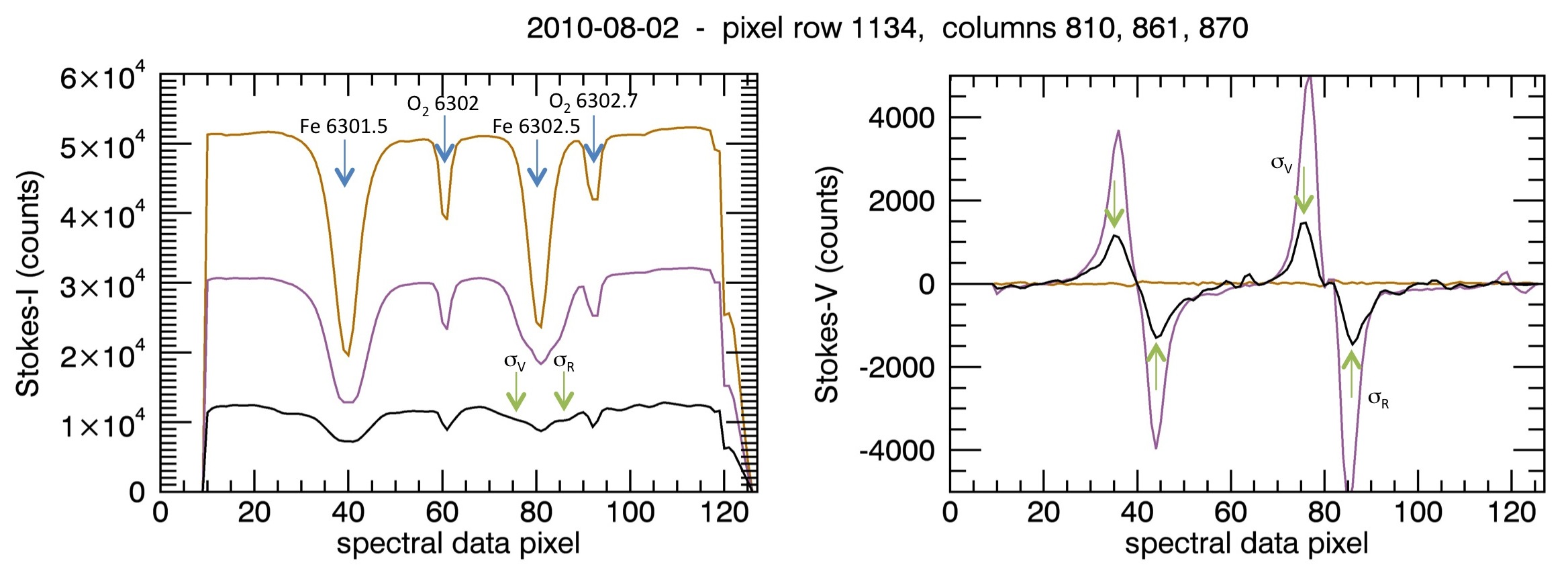}
     \caption[Example Stokes-I and Stokes-V profiles]{Examples of Stokes-I (left panel) and Stoke-V (right panel) spectral profiles for three pixels across an active region.  Orange corresponds to a quiet-sun pixel, purple to a penumbral pixel, and black to an umbral pixel.}
     \label{FIG_profiledata}
     \end{center}
     \end{figure}

In either case, the bulk processing of the code is on a  per-output-pixel basis, seeing only a single set of Stokes-I (and Stokes-V if that fit is so desired) profiles, such as those depicted in Figure~\ref{FIG_profiledata}.
The VSM-6302-vector spectra include two Zeeman-sensitive solar Fe lines at 6301.5 and 6302.5\AA, as well as two telluric O$_2$ lines at 6302.0 and 6302.8\AA.  While the 6301.5\AA-Fe-line is an anomalous-Zeeman line, the 6302.5\AA-line is not, and the triplet nature of the line can be perceived in Stokes-I in high-field-strength penumbral and umbral regions.  In regions with a strong longitudinal field, the Stokes-V profile shows clear signals corresponding to the $\sigma$-components of the triplet lines.

\begin{figure}
     \begin{center}
     \includegraphics[width=\textwidth]{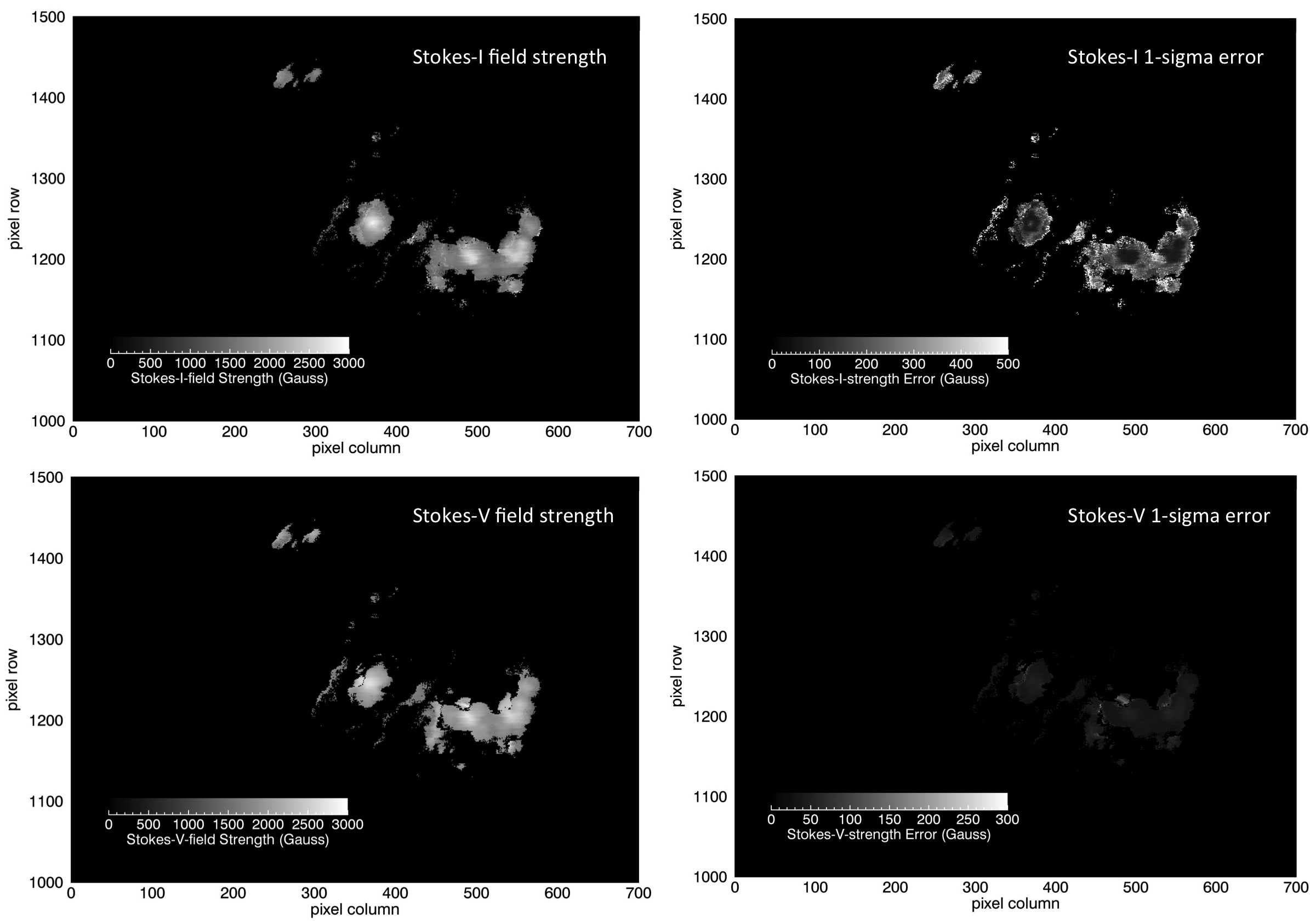}
     \caption[Example Zeemanfit results across an active region]{Results of the Zeemanfit computed output zoomed in to one region for data taken on January 11$^\mathrm{th}$, 2013.}
     \label{FIG_tiledresults}
     \end{center}
     \end{figure}

\begin{figure}
     \begin{center}
     \includegraphics[scale=0.16]{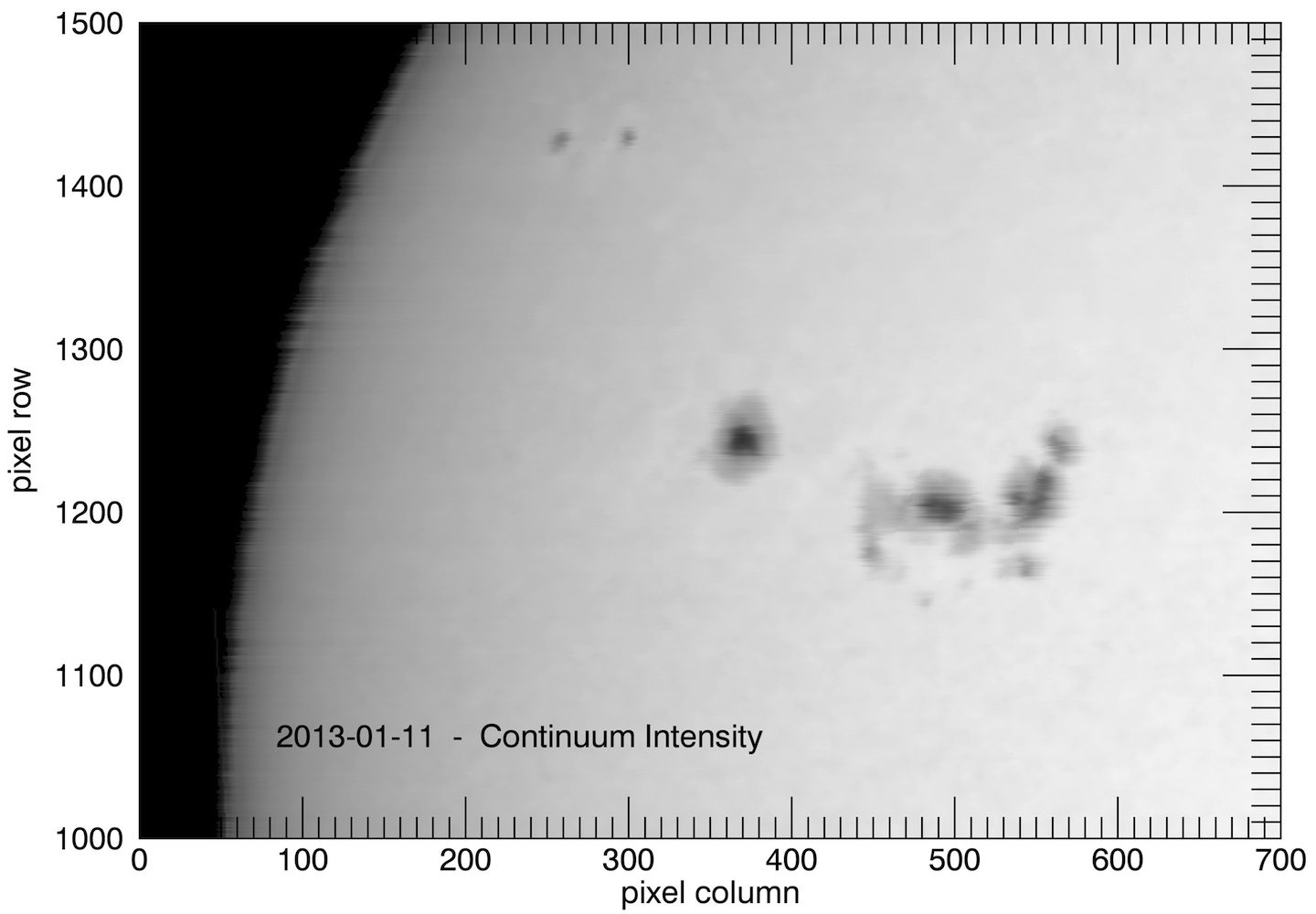}
     \caption[Continuum image]{Continuum image corresponding to Figure~\ref{FIG_tiledresults} and the January 11$^\mathrm{th}$, 2013 data.}
     \label{FIG_continuumdata}
     \end{center}
     \end{figure}

The calculation of the magnetic-field strength using the Zeemanfit code is as simple as it can be (\S\ref{Code_Calc}), relying on only the line-center separation of the two 6302.5\AA-$\sigma$-component lines fit to the profile.  Figure~\ref{FIG_tiledresults}
depicts the resultant output for one zoomed-in region of one data image, while Figure~\ref{FIG_continuumdata}
shows a continuum-reference map of that region.  The computed output in this example includes:
\begin{enumerate}
\item The total magnetic-field strength calculated from the fit to the Stokes-I profile.
\item The Stokes-I 1-$\sigma$ error propagated from the fit-uncertainty in the line-center positions.
\item The longitudinal magnetic-field strength calculated from the fit to the Stokes-V profile.
\item The Stokes-V 1-$\sigma$ error propagated from the fit-uncertainty in the line-center positions.
\end{enumerate}
Note that not all of the pixels fit in Stokes-I have corresponding measurements from Stokes-V.  This is primarily because the Stokes-V signal may become too weak to fit in the presence of a strongly transverse magnetic field.  The results of fitting the Stokes-V profiles are not standard-pipleline data products.

For the standard products, one feature of the Zeemanfit Stokes-I calculation is that it is designed around the greatest ease of fitting the highest-field-strength pixels.  Therefore, the uncertainties in the Stokes-I measurements tend to be higher in the lower-strength outer regions of the penumbrae.

Also outside of standard pipeline data production, if the flag /DISPERCALC is set when the Zeemanfit routine is called, then the dispersion is calculated on a per-output-pixel basis for {\em all} output pixels using the fit positions of the two telluric O$_2$ lines (see \S\ref{Code_Disper}), and two more frames are added to the output fits file, one containing the calculated dispersion in \AA/pixel, the other containing the relative 1-$\sigma$ error in \% on the dispersion measurement propagated from the uncertainty in the fit line-center positions of the telluric lines.  This option does add significantly to the processing time, as all on-disk pixel-spectra must be fit in order to report the measured dispersion across the entire image.

\section[\textcolor{red}{Pipeline Version Changes}]{\textcolor{red}{Pipeline Version Changes}}
\label{Versions}

The earliest Zeemanfit results output by the SOLIS pipeline (6302v results only) have header-keyword values PROVER2A = '1.1' and PROVER2B = '1.1'.  The following adjustments affecting the 6302v-Zeemanfit processing have since been implemented for \textcolor{red}{PROVER2A = '1.2' and PROVER2B = '1.2'}:
\begin{enumerate}
\item The use of the {\bf solis\_vsm\_zeemanfit\_FITRANGE()} function has replaced the previously hard-coded values of fitrange = [15,105].  These boundaries (inclusive) define the range of spectral-data pixels that are passed to MPFITFUN() to fit the 6302 input profiles.  See \S\ref{Code_Layout} for further information about solis\_vsm\_zeemanfit\_FITRANGE().
\item {\bf The method used to guess the Fe $\pi$-component line-center positions} for the Stokes-I Zeeman-fit has been modified to better account for possible variations in particularly the symmetry/asymmetry of the Zeeman-triplet profiles.  See \S\ref{Code_Fit_I} for details.
\item Limits have been imposed such that now {\em none} of the line profiles used in the Stokes-I Zeeman-fit may be defined in emission; {\bf now all Stokes-I Zeemanfit line-amplitudes must be less than zero}.  Previously, the three lines used to fit the 6301.5\AA\, Zeeman-broadened line-set had no constraints on any of their line parameters.  See \S\ref{Code_Fit_I} for details of the Stokes-I Zeemanfit setup.
\end{enumerate}

\begin{figure}
     \begin{center}
     \includegraphics[width=\textwidth]{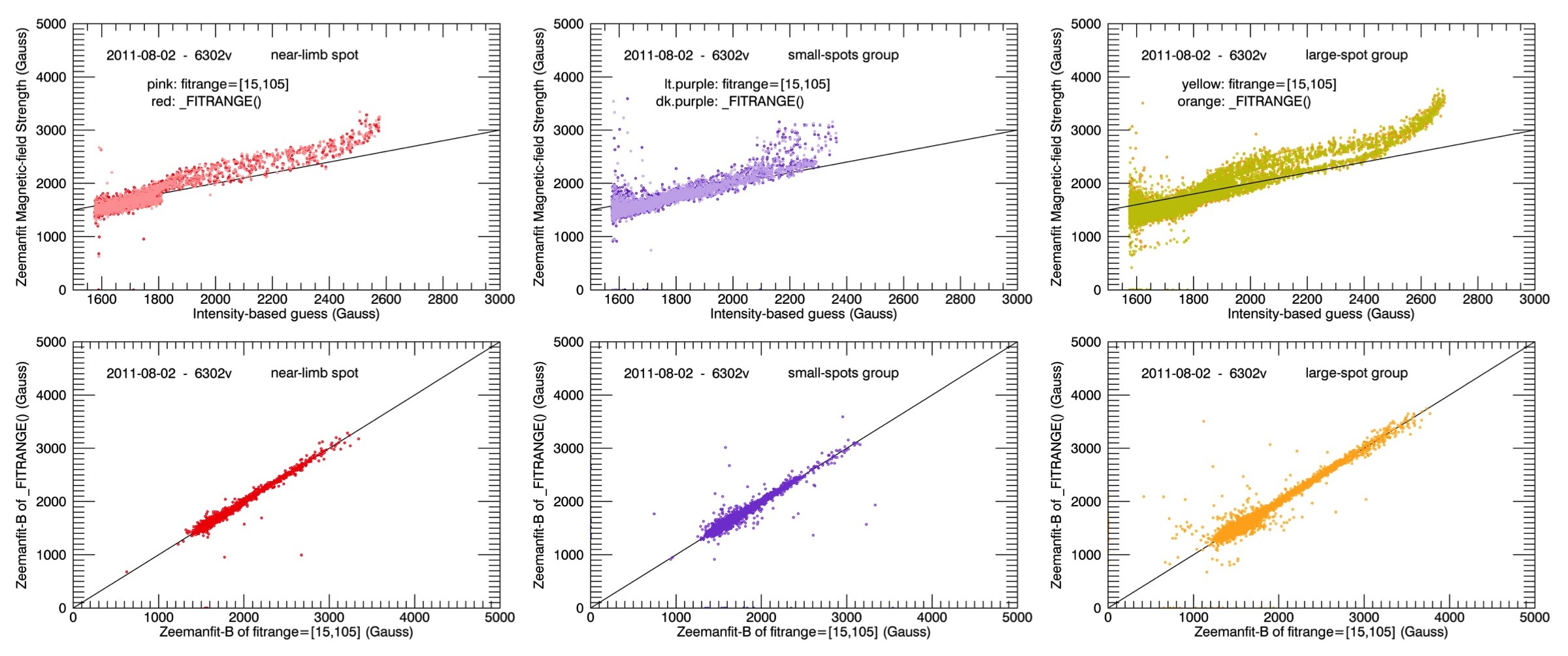}
     \caption[Scatter plots comparing 6302v results with the implementation of \_FITRANGE()]{Plots comparing the Zeemanfit results of processing a 6302v observation before and after implementation of the \_FITRANGE() algorithm.  This observation has three main sunspot groups.  The top row of plots overlay the profiles of the two run-fits as a function the intensity-guessed magnetic-field-strength for each sunspot group.  The bottom row of plots compares the two Zeemanfit results point-to-point with the old-style along the x-axis and the new-style along the y-axis.}
     \label{FIG_compareNowFITRANGE}
     \end{center}
     \end{figure}

\begin{figure}
     \begin{center}
     \includegraphics[width=\textwidth]{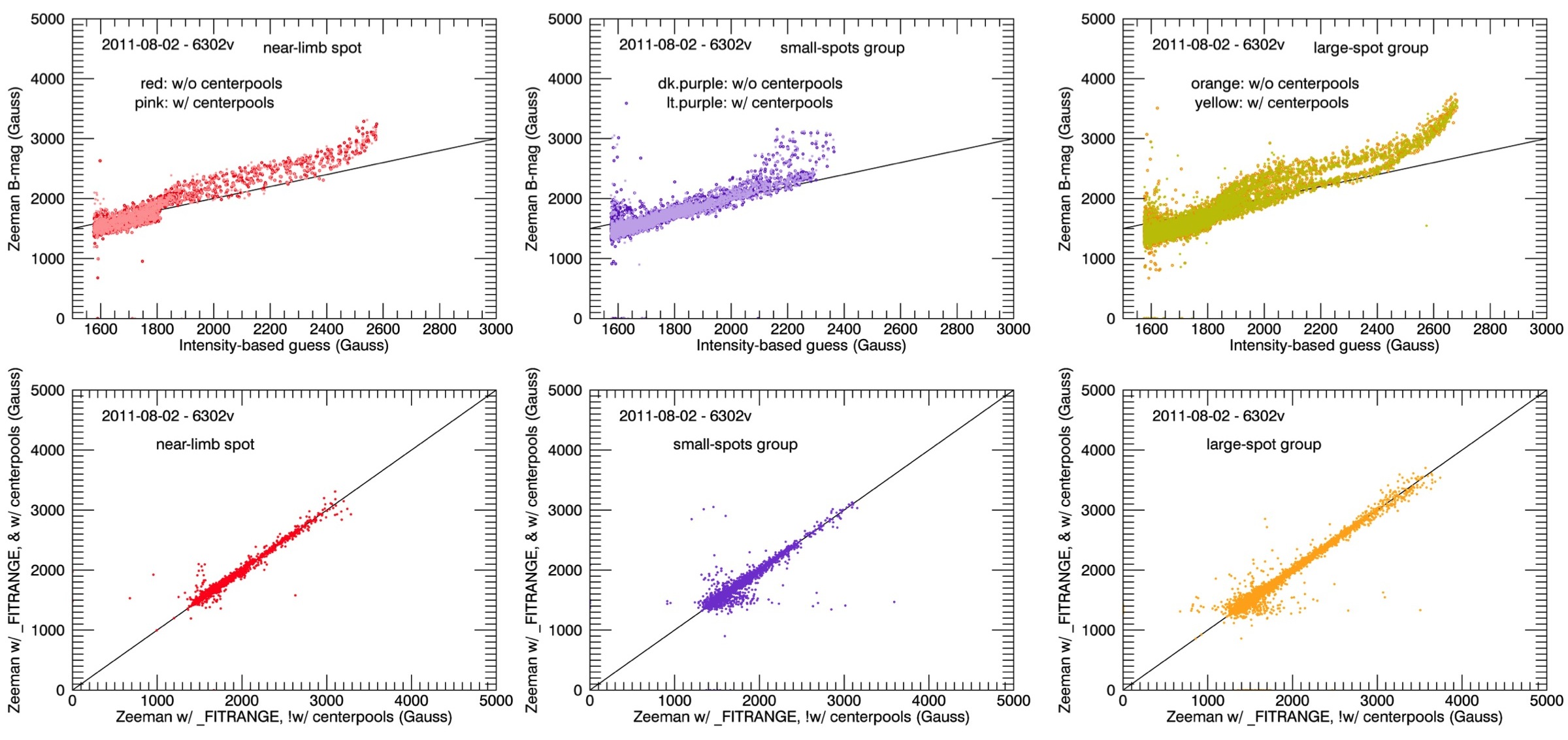}
     \caption[Scatter plots comparing 6302v results with the implementation of center pools guessing]{Plots comparing the Zeemanfit results of processing a 6302v observation before and after implementation of centerpools-style guessing of the $\pi$-component positions.  (Both runs compared here had \_FITRANGE() already implemented.)  These plots are for runs using the same observation as in Figure~\ref{FIG_compareNowFITRANGE}, and the top and bottom rows of plots follow the same conventions as in Figure~\ref{FIG_compareNowFITRANGE}.}
     \label{FIG_compareNowCenterpools}
     \end{center}
     \end{figure}

These changes were implemented into the code in the order they have been listed here.  For changes 1 and 2, Figures~\ref{FIG_compareNowFITRANGE} and Figure~\ref{FIG_compareNowCenterpools}
provide a comparison of before and after Zeemanfit results using a 6302v observation taken on August 2, 2011.  Note that for the bottom row of plots in Figure~\ref{FIG_compareNowCenterpools} (the change to using centerpools $\pi$-position guessing), each sunspot group shows a small cluster of points (above the line for the near-limb spot and below the line for the others) with a spread in results along one axis and positioned at $\sim1500$ Gauss on the other.  For these, the below-line clusters likely represent improved fit stability for those points, whereas the above-line cluster likely represents some cases of increased instability.  Zeemanfit is, unfortunately, fairly sensitive to the input initial conditions, as detailed throughout \S\ref{Code}.

\section[Operation of the Code]{Operation of the Code}
\label{Code}

The following sections detail the specifics of the code used to make the Zeemanfit--Magnetic-field-strength calculations, including: 
\begin{list}{}{}
     \item \S\ref{Code_Layout}:  The layout and functional organization of the code.
     \item \S\ref{Code_IO}:  Considerations specifically focused on the input/output of the data/results.
     \item \S\ref{Code_Sorting}: The particulars of selecting which image pixels to Zeeman-fit.
     \item \S\ref{Code_Fit}:  The particulars of the choices made for fitting the spectral profiles.
     \item \S\ref{Code_TripletPos}: Issues and resolutions concerning the fit separation of the model-Zeeman triplet lines.
     \item \S\ref{Code_Disper}:  The dispersion used/calculated by the code.
     \item \S\ref{Code_Calc}:  The particulars of the magnetic-field-strength calculation and error-output.
\end{list}

     \subsection[Layout and Function Descriptions]{Layout and Function Descriptions}
     \label{Code_Layout}

The code is currently written in idl, using READFITS and WRITEFITS to ingest spectral data and output the zeemanfit results.  The primary foundation for the choice of idl is the use of MPFITFUN as the routine used for fitting a line-profile to each image pixel of spectral data.  Any translation out of idl would need to begin with either a new fit-finder routine chosen in the new language, or setting up the new-language-code to take MPFITFUN as an external idl routine.

The hierarchy of code usage is as follows (functions placed more than once are written in grey):
\begin{list}{}{}
\item {\bf solis\_vsm\_ZEEMANFIT()}
	\begin{list}{{\bf*}}{}
	\item solis\_vsm\_zeemanfit\_IO()
	\item solis\_vsm\_zeemanfit\_STOKESVPICK()
		\begin{list}{$\bullet$}{}
		\item solis\_vsm\_zeemanfit\_STOKESVPICK\_WIDE()
		\end{list}
	\item solis\_vsm\_zeemanfit\_DARKPIXELS()
	\item solis\_vsm\_zeemanfit\_RECIPE()
		\begin{list}{$\bullet$}{}
		\item solis\_vsm\_zeemanfit\_FITRANGE()
		\item solis\_vsm\_zeemanfit\_VFIT\_6302V()
			\begin{list}{$\circ$}{}
			\item \textcolor{grey}{solis\_vsm\_zeemanfit\_MULTVOIGTFIT()
				\begin{list}{-}{}
				\item solis\_vsm\_zeemanfit\_INDEXSTRING()
				\item solis\_vsm\_zeemanfit\_MULTVOIGT()
				\end{list}}
			\end{list}
		\item solis\_vsm\_zeemanfit\_IFIT\_6302V()
			\begin{list}{$\circ$}{}
			\item \textcolor{grey}{solis\_vsm\_zeemanfit\_FLOODMIN()}
			\item solis\_vsm\_zeemanfit\_HANDLER\_I\_6302V()
				\begin{list}{-}{}
				\item \textcolor{grey}{solis\_vsm\_zeemanfit\_MULTVOIGT()}
				\end{list}
			\item solis\_vsm\_zeemanfit\_HANDLER\_I\_6302V\_AMPCAP()
				\begin{list}{-}{}
				\item \textcolor{grey}{solis\_vsm\_zeemanfit\_MULTVOIGT()}
				\end{list}
			\item \textcolor{grey}{solis\_vsm\_zeemanfit\_MULTVOIGTFIT()
				\begin{list}{-}{}
				\item solis\_vsm\_zeemanfit\_INDEXSTRING()
				\item solis\_vsm\_zeemanfit\_MULTVOIGT()
				\end{list}}
			\end{list}
		\item solis\_vsm\_zeemanfit\_OUTFIT\_6302V()
			\begin{list}{$\circ$}{}
			\item \textcolor{grey}{solis\_vsm\_zeemanfit\_FLOODMIN()}
			\item \textcolor{grey}{solis\_vsm\_zeemanfit\_MULTVOIGTFIT()
				\begin{list}{-}{}
				\item solis\_vsm\_zeemanfit\_INDEXSTRING()
				\item solis\_vsm\_zeemanfit\_MULTVOIGT()
				\end{list}}
			\end{list}
		\end{list}
	\item solis\_vsm\_zeemanfit\_HEADER()
	\item solis\_vsm\_zeemanfit\_OUTNAME()
	\end{list}
\item {\bf solis\_vsm\_zeemanfit\_PLOT()}
	\begin{list}{{\bf *}}{}
	\item \textcolor{grey}{solis\_vsm\_zeemanfit\_MULTVOIGT()}
	\end{list}
\end{list}
The individual function descriptions are given below:
\\*

\noindent {\bf SOLIS\_VSM\_ZEEMANFIT:}
The Zeemanfit code is accessed at the command line through the \nohyphens{SOLIS\_VSM\_ZEEMANFIT()} function.  This function:
\begin{list}{*}{}
     \item Calls \_ZEEMANFIT\_IO() to read in the input spectra-files and sort the data into a useful organization.
     \item Sorts and organizes the input parameters in preparation for calling \_ZEEMANFIT\_RECIPE().
     \item Calls \_ZEEMANFIT\_STOKESVPICK() and \_ZEEMANFIT\_DARKPIXELS() to select which image pixels should be Zeemanfit (see \S\ref{Code_Sorting} for more about the selection criteria).
     \item For each Zeemanfit-viable image pixel, calls \_ZEEMANFIT\_RECIPE() for Zeemanfit--magnetic-strength results.
     \item Calls \_ZEEMANFIT\_HEADER() to create an output header, and \_ZEEMANFIT\_OUTNAME() to define the output filename, and writes the primary results to a fits file.  
\end{list}
At the command line, the call returns a structure-array of results for each image pixel.  The default call settings return enough information to call the SOLIS\_VSM\_ZEEMANFIT\_PLOT() function to return line-profiles to plot up the line-fit to the spectral data of a given pixel.  Set the /TERSE flag to return only the primary results of the Zeeman fitting (still slightly more (RMSFIT) than is written to the output fits file).  Set the /VERBOSE flag to return enough information to diagnose the results relative to the initial fit-guesses.

Be sure to set the /VFIT flag if you want the results to include the longitudinal field strength computed from the fit to the Stokes-V spectra, and the /WRITEFILE flag to actually write the results to a file rather than {\em only} returning them into 'resultarray' (below) at the command line.  /QUIET turns off printed warnings and progress-markers, and several other keyword values allow for variations on, primarily, the constraints used in fitting the Stokes-I spectra, such as OGAUSS, which takes a vector of 1s and/or 0s setting specific lines in the fit to be fit only as gaussian's (or not).

For standard pipeline computing, at least {\bf two} input files must be specified:  The Stokes-I spec-cube file is taken as the primary input; PFILE specifies the regular Level-2 output file, which contains a frame for the Level-2 corrected continuum image; and VFILE specifies the Stokes-V spec-cube, which must be supplied to run /Vfit and {\em can} be used for selecting pixels and guessing input profiles if PFILE is not specified, though the results will vary measurably from those produced using PFILE.  See \S\ref{Code_IO} for more detail on the input/output and on using Level-0.5 scan-line files rather than the Level-1.5 spec-cubes.

The name of the output fits file is derived from the name of the first spectra-input file and header information.  To add a suffix of additional information to the output-file name, use the OUTSFX keyword.  To add a prefix and/or specify a directory path into which the output file should be placed, use the OUTPRFX keyword.  

For basic-pipeline processing, the call statement should  look something like:
\begin{verbatim}
    resultarray = SOLIS_VSM_ZEEMANFIT ('Data/k4v9t130207t171846_i.fts.gz', $
        PFILE='Data/k4v92130207t171846.fts.gz', $
        OUTPRFX='ResultsStorageDirectory/', $
        /WRITEFILE, /TERSE, /QUIET)
\end{verbatim}

     {\bf solis\_vsm\_zeemanfit\_DARKPIXELS:}
     This function is contained within the source-code file solis\_vsm\_zeemanfit.pro.  It is used to select all pixels across the image (using the Level-2 corrected continuum-intensity image) that are dark enough (and on-disk enough) to qualify as sunspot pixels that should be Zeeman-fit (even if Stokes-V selection and fitting are not viable).
     
     It takes in a Level-2 fits file, reading in the continuum-intensity frame as well as the parameters used to model the limb-darkened quiet-sun intensity.  It computes the limb-darkened intensity, $I_{LD}$, using the equation:
     \begin{equation}
        I_{LD}
     =
        I_0 \sum_{n=1}^6{ \mu_n \times \left[ 1  - \left(\frac{r}{R_{LD}}\right)^2 \right]^{n/2} }
     \, ,
     \label{EQ_limbdark}
     \end{equation}
where $r$ is the pixel's distance from the designated disk center, and $I_0$, $R_{LD}$, and each of the $\mu_n$'s are the limb-darkening fit parameters read in from the PFILE header.  It then compares the limb-darkened-fit to the continuum-intensity data and selects those pixels which are:
\begin{enumerate}
     \item darker than 95\% of the limb-darkened-fit intensity
     \item located within 95\% of the distance from the disk center to the limb-darkened-fit radius, $R_{LD}$.
\end{enumerate}
passing back their $I / I_{LD}$ value for use by the Zeemanfit code.
     \\*

     {\bf solis\_vsm\_zeemanfit\_STOKESVPICK:}
     This function is contained within the source-code file solis\_vsm\_zeemanfit.pro.  It is used to select pixels across the image (using the Stokes-V-spectra peak-pixel values and locations) that show a strong signal in polarized light consistent with a decently strong local-magnetic-field (longitudinal).  This selection is used to indicate pixels for which the Stokes-V spectra may be Zeeman-fit, or to select all pixels for Zeeman-fitting (Stokes-I and Stokes-V) in the case where an appropriate Level-2 PFILE is not available to run solis\_vsm\_zeemanfit\_DARKPIXELS().  For further details on the pixel-selection methods, see \S\ref{Code_Sorting}.
     \\*

     {\bf solis\_vsm\_zeemanfit\_RECIPE:}
     This function is contained within the source-code file solis\_vsm\_zeemanfit.pro.  It takes in the user-flagged choices and spectral-data profiles for a given image pixel and computes the profile fit(s) and magnetic-field strength for that pixel.  It returns a data structure that by default contains the values for the primary results, as well as the parameters used to define the fit(s).  (See \S\ref{Code_Fit} for specifics on the different fit scenarios.)
     
     If the pixel has been flagged as {\em not} Zeemanfit-viable, zeroes are returned for the magnetic-field-strength values and, if the /DISPERCALC keyword is flagged, the Stokes-I profile is fit with only four lines by calling the SOLIS\_VSM\_ZEEMANFIT\_OUTFIT\_6302V() function so that the dispersion for that pixel may be calculated from the separation of the two telluric O$_2$ lines.
     
     If the pixel {\em is} flagged as Zeemanfit-viable, the Stokes-V (if /VFIT is flagged) and Stokes-I spectra are fit separately by calling SOLIS\_VSM\_ZEEMANFIT\_VFIT\_6302V() and SOLIS\_VSM\_ZEEMANFIT\_IFIT\_6302V().  If /DISPERCALC is flagged, the dispersion is then calculated from the Stokes-I positions of the two telluric O$_2$ lines, otherwise the dispersion is taken as set for a given camera (see \S \ref{Code_Disper} for details).
     Lastly the magnetic-field strength is calculated from the separation of the two 6302.5\AA \ $\sigma$-component lines (see \S \ref{Code_Calc} for details).  If either (a) the computed field-strength is $> 5000$~Gauss, or (b) the computed 1-sigma error is $> 1000$~Gauss, the values are clipped out and returned as zeroes (though the line-fit parameters are returned to the command-line intact for potential investigation).
     \\*

{\bf solis\_vsm\_zeemanfit\_FITRANGE:}
This function is contained within the source-code file solis\_vsm\_zeemanfit.pro.  It is called from solis\_vsm\_zeemanfit\_RECIPE() (if fitrange has not already been externally set by the user), taking in an individual Stokes-I profile to return a vector of the first and last pixel indices defining the range of spectral-data pixels that should be passed to MPFITFUN() for the profile fit(s).

The algorithm does rely on a number of hard-coded values related to the input Stokes-I data profile.  It assumes that:
\begin{list}{*}{}
   \item That the data profile is measured across 128 pixels.
   \item That blue-side transmission has reached roughly continuum levels by pixel 20, while the 6301.5 Fe line-set minimum occurs somewhere between pixels 20 and 63.
   \item That no more than about two-thirds of a sometimes encroaching blue-ward telluric line occurs within 28 pixels from the minimum position of the 6301.5-Fe-line set.
   \item And that a red-side data peak occurs {\em redward} of the second telluric O$_2$ line within the pixel range 86 to 122.
\end{list}
If the above conditions are met, \_FITRANGE() is able to return bounds defined by the blue-side and red-side peaks in the continuum of the profile (there-by avoiding transmission drop-off regions), but that on the blue side exclude the presence of an additional (usually present in the 6302l spectra) telluric feature not accounted for in the Zeeman-model fit.

Please see \S\ref{Code_Fit} for a figure presenting an example map of \_FITRANGE()-selected bounds, and \S\ref{Code_Fit_I} for a figure presenting an example comparison of 6302v versus 6302l spectral profiles.
     \\*

\noindent {\bf solis\_vsm\_zeemanfit\_VFIT\_6302V:}
This function returns the fit parameters to a Stokes-V profile.  It takes in the profile data; sets up guesses for the fit, evaluating the minimum/maximum positions of the first and last halves of the V-spectra-profile for line-center guesses; and calls \nohyphens{SOLIS\_VSM\_ZEEMANFIT\_MULTVOIGTFIT()} to do the actual work of fitting multiple (4) line profiles to the Stokes-V data.  See \S \ref{Code_Fit_V} for further details.
\\*

\noindent {\bf solis\_vsm\_zeemanfit\_IFIT\_6302V:}
This function returns the fit parameters to a Stokes-I profile under the assumption that the two 6302\AA \ Fe lines are split into triplets.  It takes in the profile data; sets up guesses for the fit, using \_ZEEMANFIT\_FLOODMIN() to initialize guesses for the primary line-center positions, and guessing the $\pi$-$\sigma$ component separations based on a continuum-intensity first-estimate of the magnetic-field strength (\S\ref{Code_TripletPos}).  It then calls \_ZEEMANFIT\_MULTVOIGTFIT() to do the actual work of fitting multiple (8) line profiles to the Stokes-I data.  See \S\ref{Code_Fit_I} for further details.
\\*


     {\bf solis\_vsm\_zeemanfit\_HANDLER\_I\_6302V:}
     This function is contained within the source-code file solis\_vsm\_zeemanfit\_ifit\_6302v.pro.  It can be passed to \_ZEEMANFIT\_MULTVOIGTFIT() and used as a go-between function between MPFITFUN() and \_ZEEMANFIT\_MULTVOIGT(), allowing MPFITFUN() to vary the {\bf separation} between the 6302.5\AA \ $\sigma$- and $\pi$-components, even though the solis\_vsm\_zeemanfit\_MULTVOIGT() function requires the line-center {\bf positions} of each component.
     
     \begin{eqnarray}
       x_{0,\sigma_R} &=& x_{0,\pi} + {\bf c_{xR}}
       \, ,
     \\
       x_{0,\sigma_V} &=& x_{0,\pi} - {\bf c_{xV}}
       \, ,
     \end{eqnarray}
where $c_{xR}$ and $c_{xV}$ specify the separation of the $\pi$-component to the right and left $\sigma$-components.
 
     The $c_{xR}$ and $c_{xV}$ parameters are used (rather than setting an inequality statement in the parinfo[x0].tied parameter) to constrain the three 6302.5\AA \ triplet lines to remain in the wavelength order specified.  It also allows the error in the magnetic-field-strength calculation to depend on only the error in the $\sigma$-$\pi$ separation {\em if} the two separations are mirrored (see \S \ref{Code_TripletPos}), rather than on the error in both relevant line-center positions.
     \\*
     
     {\bf solis\_vsm\_zeemanfit\_HANDLER\_I\_6302V\_AMPCAP: }
     This function is contained within the source-code file solis\_vsm\_zeemanfit\_ifit\_6302v.pro.  It can be passed to \_ZEEMANFIT\_MULTVOIGTFIT() and used as a go-between function between MPFITFUN() and \_ZEEMANFIT\_MULTVOIGT().
     
     This handler function serves the same functionality as \_ZEEMANFIT\_HANDLER\_I\_6302V() with the addition of constraining the two 6302.5\AA \ $\sigma$-component amplitudes to be less than or equal to the $\pi$-component amplitude (so that MPFITFUN() varies an amplitude-scaling-parameter for each $\sigma$-component within the range of 0.0 to 1.0).
     
     \begin{equation}
       A_{\sigma_{R/V}} = A_\pi * {\bf c_{A,R/V} }
     \end{equation}
     \\*

\noindent {\bf solis\_vsm\_zeemanfit\_OUTFIT\_6302V:}
This function returns the fit parameters to a Stokes-I profile under the assumption that the two 6302\AA \ Fe lines are {\bf not} split into triplets.  It takes in the data profile; sets up guesses for the fit, using \_ZEEMANFIT\_FLOODMIN() to pick a guess for each of the line-center positions; then calls \nohyphens{\_ZEEMANFIT\_MULTVOIGTFIT()} to do the actual work of fitting multiple (4) line profiles to the Stokes-I data.  It is used to find the separation of the two telluric O$_2$ lines to calculate local dispersion (if requested) in the cases of quiet-sun spectra.  See \S \ref{Code_Fit_out} for further details.
\\*

\noindent {\bf solis\_vsm\_zeemanfit\_MULTVOIGT:}
This function takes in a vector of independent values plus a parameter array of Voigt-profile line parameters for multiple lines, and returns the calculated line profile.  The function assumes that the last row of the parameter array contains coefficients for a polynomial background continuum.  It takes in Voigt-profile parameters for each row of the parameter-array in the order: Amplitude, line-center position, Gauss-width $\sigma$, Lorentz-width $\gamma$.  If either of the two width parameters are set to zero, it calculates the profile for that line using a pure gaussian or pure Lorentzian profile, as appropriate.

The preferred format for the parameter-array is an actual {\bf array}, however, MPFITFUN() requires a {\bf vector} of parameters, therefore \_MULTVOIGT() {\bf will} take a vector of parameters under the assumption that an additional final element has been added specifying the {\bf order} of the background polynomial ($n_\mathrm{coefficients} - 1$) and that the number of elements allotted per line component corresponds to the maximum of 4-Voigt-parameters vs. $n_\mathrm{coefficients}$-of-the-polynomial.

Flagging the keyword COMPONENTS=components will return an additional array of line-profile values corresponding to each continuum+spectral-line component of the full profile.
\\*

\noindent {\bf solis\_vsm\_zeemanfit\_MULTVOIGTFIT:}  This function does all of the work of formatting passed-in multiple Voigt-profile--parameter guesses, accounting for everything necessary to implement all of the bells and whistles and requirements of fitting them, and converting them into the format used by MPFITFUN().

Using the PARINFO\_ADD=parinfo\_add keyword, it can take in a structure-array of pre-set constraints for running MPFITFUN().  However, {\bf it} creates the appropriate parinfo-structure-vector to be passed to MPFITFUN() that is set up for running a multiple-Voigt-profiles--plus--polynomial-background--fit to the data using the model in the \_ZEEMANFIT\_MULTVOIGT() function.
\begin{list}{*}{}
     \item If the OGAUS or OLORN keyword parameters are included in the call, it will fix parinfo constraints so that the specified lines are fit as gaussians or Lorentzians only.
     \item It also reformats the parameter-(guess)-array into a parameter vector so that MPFITFUN() can handle it properly, then converts it back into an array to return as the results to the user call.  It calls \_ZEEMANFIT\_INDEXSTRING() to convert any specific parameter indexes named in the PARINFO\_ADD[].tied values to their corresponding vector specifications.
     \item It does not use any .value specifications passed through the PARINFO\_ADD keyword, only the guesses in the passed param-guess array or their appropriate translations.
\end{list}

In order to allow MPFITFUN() to run its fit on alternative parameters to those explicitly accepted by the \_ZEEMANFIT\_MULTVOIGT() function, using the HANDLER keyword, \_ZEEMANFIT\_MULTVOIGTFIT() can take in the name of a handler function to pass to MPFITFUN() in lieu of \_ZEEMANFIT\_MULTVOIGT() {\bf if} the handler function conforms to the following properties:
\begin{enumerate}
     \item Conforms to the MPFITFUN() function requirements of returning a line-profile vector to a call statement that looks like:
         \begin{verbatim}
         lineprofile = H_FUNCTION(x_vector, param_vector)
         \end{verbatim}
     (Assuming that param\_vector is formatted appropriate to MPFITFUN() but passing along to \_ZEEMANFIT\_MULTVOIGT() the type of param\_vector that {\bf it} expects, i.e., 'handling' the passed parameters.)
     \item Alternatively, returns an MPFITUN()-suitable converted parinfo-structure-vector \\ (e.g., parinfo[$x_{0,\sigma_R}$].value $\rightarrow$ parinfo[$c_{xR}$].value) to a call statement if the /PARINFO\_CONVERT flag is set:
         \begin{verbatim}
      parinfo_vector = H_FUNCTION(0, parinfo_vector, /PARINFO_CONVERT)
         \end{verbatim}
     \item Alternatively, returns a reverse-converted parameter-vector (e.g., $x_{0,\sigma_R} \leftarrow c_{xR}$) to a call statement if the /RCONVERT flag is set:
         \begin{verbatim}
        param_vector = H_FUNCTION(0, param_vector, /RCONVERT)
         \end{verbatim}
\end{enumerate}

This function (\_ZEEMANFIT\_MULTVOIGTFIT() ) will also return PERROR, BESTNORM, and DOF as calculated by MPFITFUN() if those keywords are set in the call.
\\*

     {\bf solis\_vsm\_zeemanfit\_INDEXSTRING:}
     This function is contained within the source-code file solis\_vsm\_zeemanfit\_multvoigtfit.pro.  It takes in a string variable potentially containing an array designation, e.g., 'P[5,2]', + a vector of array dimensions, and returns a string containing the vector-translated designation, e.g., 'P[17]'.
     \\*

%

\noindent \textcolor{blue}{{\bf solis\_vsm\_zeemanfit\_FLOODMIN:}}
This function is used by both solis\_vsm\_ZEEMANFIT\_IFIT\_6302V() and solis\_vsm\_ZEEMANFIT\_OUTFIT\_6302V() to locate a specified number of data-profile minimums within a specified data-range (i.e., to find the four primary minima corresponding to the two solar Fe lines and two telluric O$_2$ lines).

This minimum finder works by evaluating the number of data {\em ranges} that fall below some specified floodline value.  It begins with a floodline set to the minimum value of the input data, then moves upward in increments (the number of which is optionally adjustable) toward the data maximum until the number of minima pools matches the number of minima requested.  It returns -1 if this criterion is not met.  In order to avoid returning local 'minima' due only to background noise, a minimum gap value may be set specifying the required pixel distance between minima pools.  However, this function is still best used only with relatively smoothly varying data.

Note that a search-data-range must be specified in order to avoid returning data-boundary positions where the transmission may drop toward zero.
\\*

\noindent \textcolor{blue}{{\bf solis\_vsm\_zeemanfit\_IO:}}
This function is used by SOLIS\_VSM\_ZEEMANFIT() to take in and sort the input data files.  It returns spec-cubes of the the Stokes-I and Stokes-V (if available) spectral data, as well as a continuum-intensity frame to write to the output file if an appropriate PFILE has been specified.  There are currently a number of different data-input configuration options.  See \S\ref{Code_IO} for more details.
\\*

\noindent \textcolor{blue}{{\bf solis\_vsm\_zeemanfit\_HEADER():}}
This function is called by SOLIS\_VSM\_ZEEMANFIT() to create an appropriate fits header for the fits output file.  It consults SOLIS\_VSM\_FITSINFO() for frame specifications to describe the appropriate frames output in accordance with the spectrally-derived-quantities data-product file specifications.
\\*

\noindent {\bf solis\_vsm\_zeemanfit\_OUTNAME():}
This function is called by SOLIS\_VSM\_ZEEMANFIT() and uses the primary input filename and header to define an output filename that corresponds to the current SOLIS file-naming conventions.
\\*

\noindent {\bf solis\_vsm\_zeemanfit\_PLOTS:}
This function is largely independent of the rest of the solis\_vsm\_zeemanfit code.  It takes in a data structure like those returned by SOLIS\_VSM\_ZEEMANFIT\_RECIPE() (fit-results for a single image pixel), retrieves the spectral-profile used to calculate the fit, calls SOLIS\_VSM\_ZEEMANFIT\_MULTVOIGT() to evaluate the fit parameters, and returns an array containing: the wavelength axis, the data profile, the fit profile, the fit-continuum profile, and the fit-continuum+line profile for each of the fit lines.
\\*

     \subsection[Input/Output Considerations]{Input/Output Considerations}
     \label{Code_IO}

Right now the code is set up to take in either:
\begin{enumerate}
    \item spec-cube files of Level-2 geometric-distortion-corrected Stokes-I and-V (specified with VFILE=) spectra, or
    \item a Level-0.5 scan-line file or a text-file-list of scan-line files.
\end{enumerate}
The code is optimized for using the distortion-corrected spec-cube file(s) along with the PFILE=-designated corresponding Level-2 standard-output file (e.g., a Level-2 Quicklook file).  The PFILE is used to supply an image frame of the continuum intensity and, when present, is the preferred (pipeline) method for selecting which pixels should be Zeemanfit.  A Level-1 PFILE may be supplied when calculating the Zeemanfit results of assembled scan-line files, but it will not be used to select pixels, only included as a helpful-reference frame with the output fits file.  If the primary input is a Stokes-I spec-cube file and /Vfit is not flagged then one of either VFILE or an appropriate PFILE {\bf must} be supplied in order to run image-pixel selection (PFILE is preferred for default pipeline operation).  To run with /Vfit, VFILE is mandatory, while PFILE remains preferred for pixel selection.

Using the keywords XMIN, XMAX, YMIN, YMAX will tell the code to run its calculations on an area-scan-style region of the image.  Using XMIN or YMIN by themselves will tell the code to run on a single image row or image column only.  Alternatively, if using scan-line files as input, XMIN and XMAX function as described, but YMAX is disabled, and YMIN should be used to specify the row-designation (in idl 0-indexed coordinates) of the first input scan-line file.

An output fits file is only written if the /WRITEFILE keyword is flagged.  Using the OUTPRFX keyword, the call can specify a directory into which the output fits files should be written (and/or a prefix to add to the beginning of the output filename).

The names for the output files and their headers are tailored to the standard-pipeline processing and naming scheme; be careful to use the OUTSFX keyword if running multiple calculations (resulting in hopefully multiple output files) on the same data set.

If the flag /VFIT is set, then the output file will include frames for the Stokes-V calculated (longitudinal) magnetic-field strength and corresponding 1-sigma errors.  If the /DISPERCALC flag is set, then the output file will include frames for the calculated-per-image-pixel dispersion in \AA/pixel and the associated {\em relative} 1-sigma dispersion error (in \%).

     \subsection[Selecting Image Pixels to Fit]{Selecting Image Pixels to Fit}
     \label{Code_Sorting}

The code allows for two methods to select which image-pixels should be Zeeman-fit to return calculated magnetic-field strengths and errors.  The default selection method (in solis\_vsm\_ZEEMANFIT\_DARKPIXELS() ) uses the corresponding Level-2-corrected continuum intensity image to select image pixels that are noticeably darker than the background quiet sun.  The secondary method (in solis\_vsm\_ZEEMANFIT\_STOKESVPICK() ) examines the Stokes-V spectral profile for evidence of strong magnetic fields.  The appropriate pixel-selection method is called from and the results sorted within the main-level code solis\_vsm\_ZEEMANFIT().

\begin{figure}
     \begin{center}
     \includegraphics[scale=0.19]{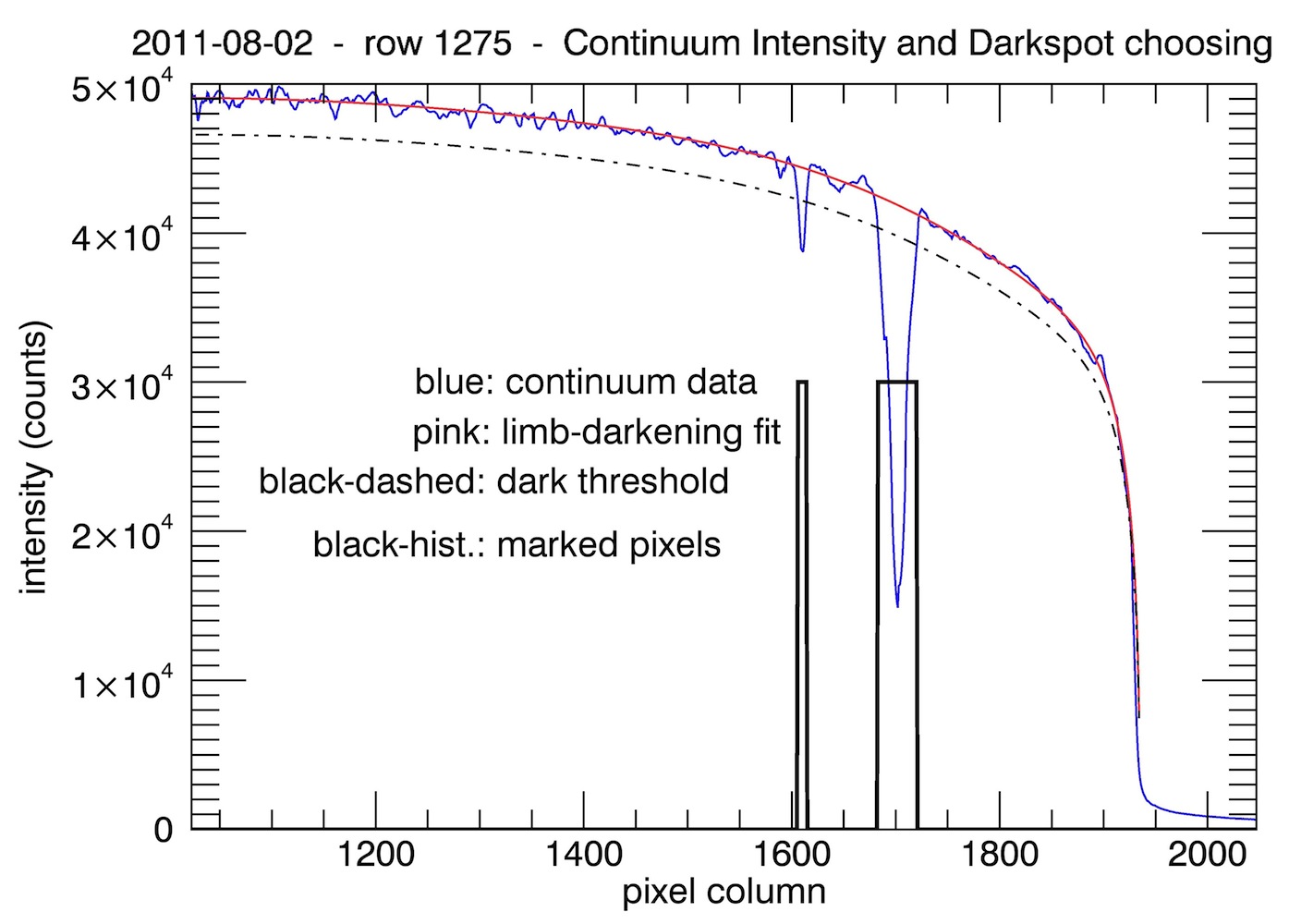}
     \caption[Choosing continuum-dark pixels for Zeeman-fitting]{Presents the continuum-intensity across a row of image-data (blue) as well as the limb-darkened-modeled fit (pink) used by solis\_vsm\_zeemanfit\_DARKPIXELS().  The threshold cutoff for selecting a pixel as dark enough to indicate the presence of a strong magnetic field is set at 95\% of the limb-darkened-fit intensity.}
     \label{FIG_limbfitfinder}
     \end{center}
     \end{figure}

Using solis\_vsm\_ZEEMANFIT\_DARKPIXELS() to select image pixels that are dark in the continuum is the preferred, default selection method because it is both more contained to the most-active regions of interest than the solis\_vsm\_ZEEMANFIT\_STOKESVPICK() selections and indiscriminate to local field orientation.  Solis\_vsm\_ZEEMANFIT\_DARKPIXELS() uses the Level-2-corrected continuum-intensity image paired with the 6-term polynomial, limb-darkening-model fit to the intensity profile to select pixels that are substantially darker than those of the quiet surroundings, as demonstrated in Figure~\ref{FIG_limbfitfinder}.

\begin{figure}
     	\begin{center}
     	\includegraphics[width=\textwidth]{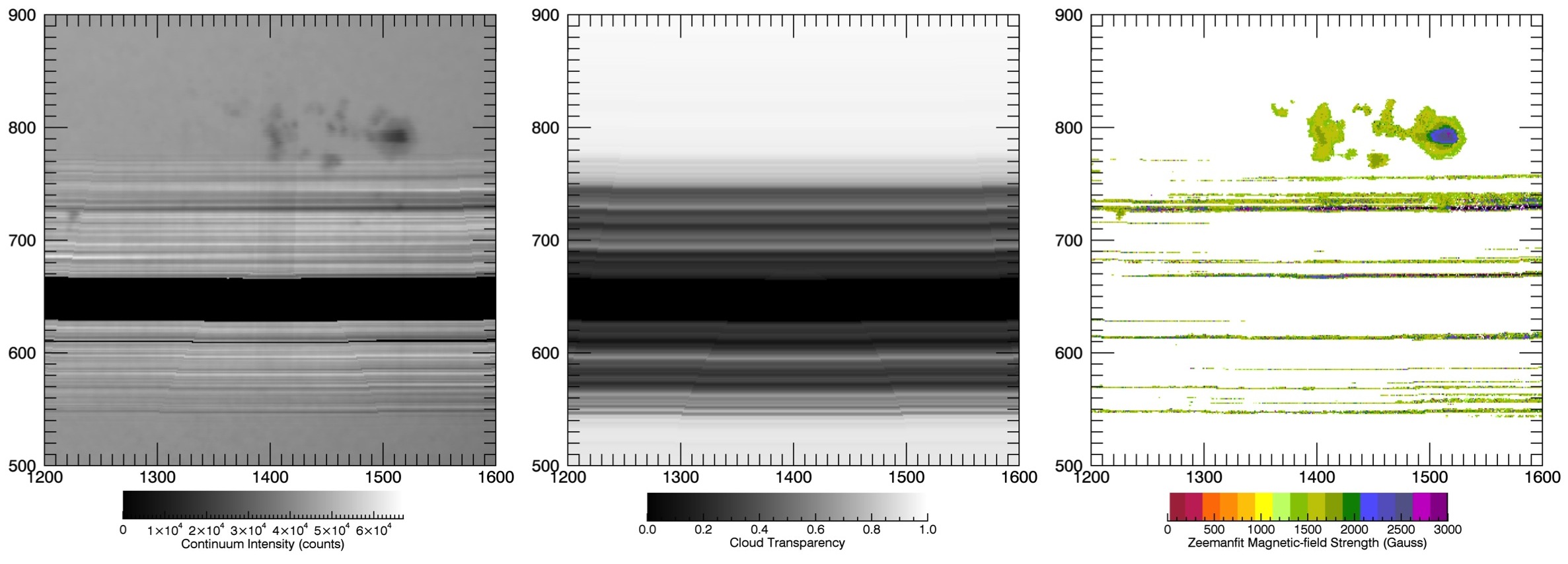}
	\caption[Zeemanfit on badly cloudy data]{6302v - 2013-07-09 - Level-2 continuum image strongly affected by clouds (left), the corresponding cloud-transparancy frame (middle), and the Zeemanfit-of-Stokes-I results if the cloud-transparency frame is ignored when selecting pixels to Zeeman-fit (right).}
     	\label{FIG_cloudstreaks}
    	\end{center}
    	\end{figure}

The primary drawback to this selection method is demonstrated in Figure~\ref{FIG_cloudstreaks}.
In the case when an observation is taken under somewhat cloudy conditions, the Level-2 continuum-intensity image cannot be perfectly normalized, leaving light or dark bands where scan lines were strongly affected by clouds.  In this case, selecting pixels-to-fit purely by their darkness relative to the limb-darkening-fit profile will include a large number of quiet-sun pixels, producing cloud-streak artifacts in the final Zeemanfit magnetogram.  Therefore, solis\_vsm\_ZEEMANFIT\_DARKPIXELS() also uses the Level-2 cloud-transparency frame to screen out pixels associated with cloudy scanlines.  Pixels are de-selected if the corresponding cloud-transparency is less than 0.9.  While this threshold may seem high, one should note that:
\begin{list}{}{}
     \item a) A threshold of 0.88 used on the observation displayed in Figure~\ref{FIG_cloudstreaks} failed to remove all cloud-streak artifacts for that observation.
     \item b) Zeemanfit is decidedly sensitive to the initial guess for the local magnetic-field-strength (as discussed in \S\ref{Code_TripletPos}), which the default operation of the code estimates directly from the relative-pixel-darkness outputted by solis\_vsm\_ZEEMANFIT\_DARKPIXELS().  Therefore, if the estimated relative-pixel-darkness is not trustworthy due to an overly-cloudy scan-line observation, the computed Zeemanfit-Stokes-I magnetic-field strength will be likewise inconsistent with the nominal calculation.
\end{list}

If the Level-2 continuum intensity image is not available, either because it was not supplied or because the Zeemanfit code is being applied to scan-line spectra that have not been positionally configured to match to Level-2 data, then pixels are selected to be fit using solis\_vsm\_ZEEMANFIT\_STOKESVPICK().  Under this method, image pixels are selected using two straight-forward criteria:
\begin{enumerate}
     \item The peak minimum {\em and} maximum values of the Stokes-V profile must be $\geq 500$ counts ($\geq 400$ counts in the case of 6302l spectra), indicating that there is a substantial amount of polarized light at that pixel.
     \item The separation of the min/max peaks for either the 6301.5\AA \ or 6302.5\AA \ lines must be $\geq 7$ pixels (indicating a reasonably high (longitudinal) magnetic-field strength).
\end{enumerate}

\begin{figure}
     \begin{center}
     \includegraphics[scale=0.6]{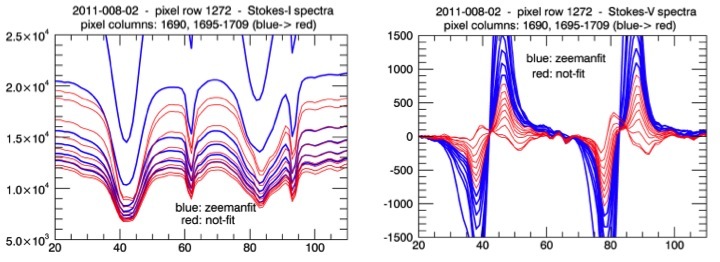}
     \caption[Stokes-I and -V profiles across a near-limb active region]{Stokes-I spectra (left panel) and Stokes-V spectra (right panel) sorted using solis\_vsm\_zeemanfit\_STOKESVPICK() for pixels in a line across a strong active region near the solar limb.  Using a slightly stricter criterion that Stokes-V selection requires a minimum or maximum amplitude of $\ge 1000$ counts, {\bf blue} indicates pixels that were selected for Zeeman-fitting, and {\bf red} indicates pixels that were not.}
     \label{FIG_notpickedprofiles}
     \end{center}
     \end{figure}

\begin{figure}
     \begin{center}
     \includegraphics[scale=0.185]{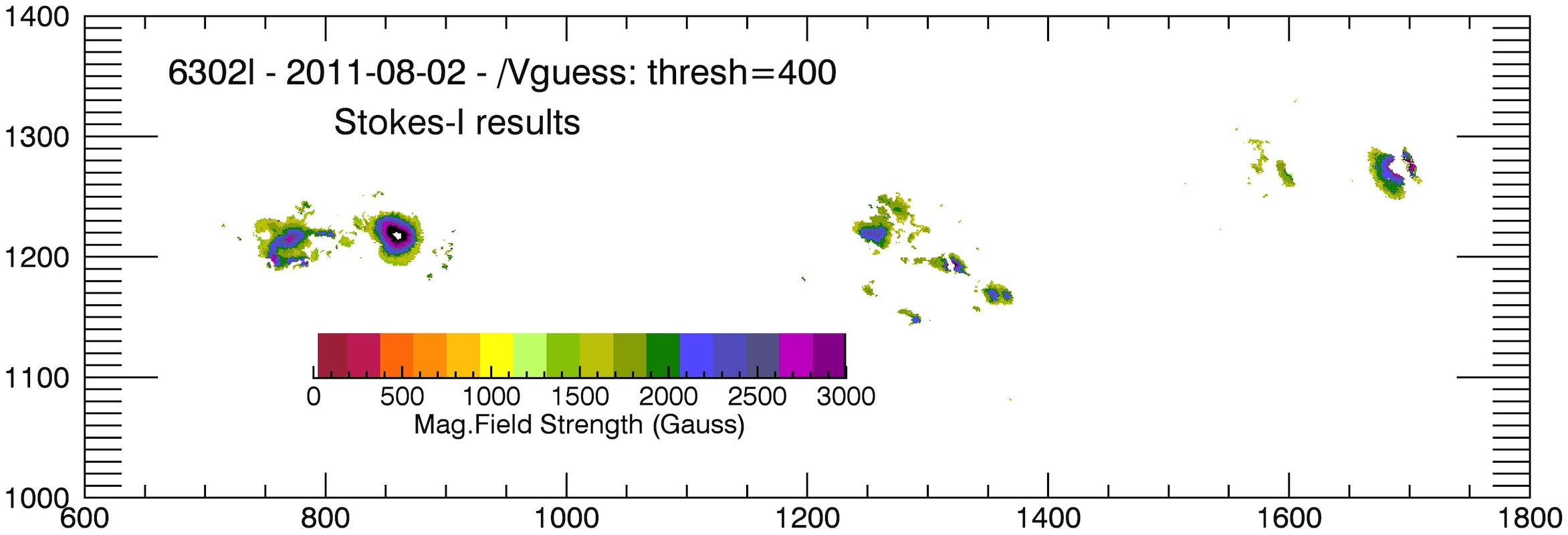}
     \caption[6302l mapped results using \_STOKESVPICK()]{Zeemanfit results for a 6302l observation where the pixels-to-fit were selected using \_STOKESVPICK() and the initial guesses for the magnetic-field strengths were supplied by the Stokes-V spectra as well.}
     \label{FIG_6302lstokesvpickMap}
     \end{center}
     \end{figure}

While the Stokes-V method provides a natural and efficient method for pixels-to-fit and line-center selection, the Stokes-V profile is not a uniformly reliable method to locate pixels of high magnetic-field strength.  Figure~\ref{FIG_notpickedprofiles}
presents Stokes-I and -V profiles for pixels in an active region that were (blue) and were not (red) selected for Zeeman-fitting using the Stokes-V selection method.  Because this active region is near the solar limb, the magnetic-field is more likely to be transverse, presenting a weak Stokes-V signal for many strong-magnetic-field pixels.  Further, even well-on-disk umbrae can have trouble with \_STOKESVPICK() if they are very dark, thereby naturally limiting the number of counts for those spectra.  This issue can be seen in the map in Figure~\ref{FIG_6302lstokesvpickMap}
of Zeemanfit results using \_STOKESVPICK() selection for some 6302l data.

     \subsection[Line-fit Choices]{Line-fit Choices}
     \label{Code_Fit}

There are three separate line-fit cases considered by this code:  Zeeman-triplet-fit to the Stokes-I spectra, Zeeman-fit to the Stokes-V spectra, and Non-zeeman-fit to the Stokes-I spectra.  These fits are orchestrated by the functions solis\_vsm\_zeemanfit\_IFIT\_6302V(), solis\_vsm\_zeemanfit\_VFIT\_6302V(), and solis\_vsm\_zeemanfit\_OUTFIT\_6302V(), respectively.

The default settings for these fit-scenarios were chosen during testing of the code to provide the most robust and efficient fit performance.  Those settings and the reasoning behind them are detailed in the sub-sections that follow.

\begin{figure}
     \begin{center}
     \includegraphics[scale=0.19]{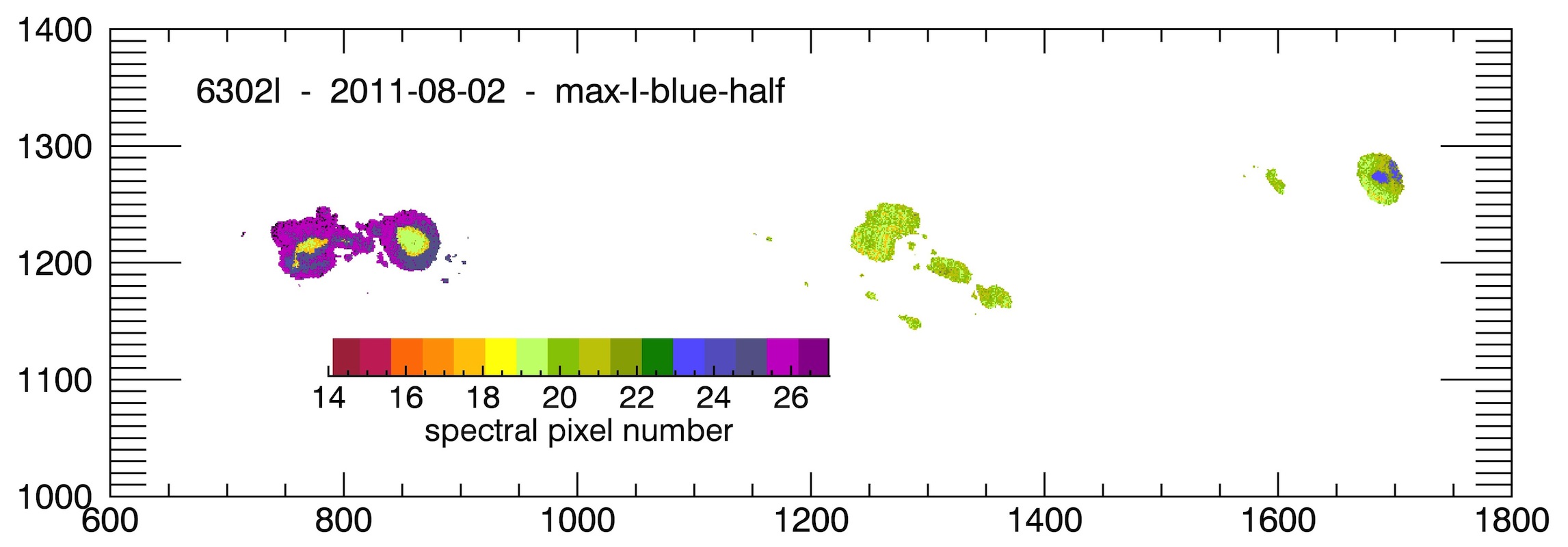}
     \caption[6302l map of blue-side pixel boundaries returned from \_FITRANGE()]{A map showing the blue-side pixel boundaries returned by \_FITRANGE() for an example 6302l observation.  Across this map, there is a total range of 13 pixels selected for this boundary, compared to a range of 9 pixels for the red-side bounds.  For a same day 6302v observation there is also a blue-boundary range of 13 pixels, but shifted left by 3 pixels, and a red-bound range of 16 pixels (though the vast majority fall into the same-range-shifted pattern as with the blue-boundary).}
     \label{FIG_6302lblueFitrange}
     \end{center}
     \end{figure}

Unless otherwise specified at the command line, the range of data-pixels used in each fit is determined on a per-image-pixel basis by a call to solis\_vsm\_zeemanfit\_FITRANGE() using that pixel's Stokes-I profile.  In an older version of the code (PROVER2A = 1.1), the fit-range was hardcoded as [15:105] (out of a total [0:127] spectral data pixels).  That range was chosen to conform to currently tested 6302v data, but was later found to be inappropriate for 6302l data.  Now, as can be seen in Figure~\ref{FIG_6302lblueFitrange},
the range of data pixels passed to MPFITFUN() does vary considerably across different regions.  However, this variation was found to be beneficial for the 6302l processing to allow for the best overall stability of the fits.


          \subsubsection[Stokes-I fit]{Stokes-I Zeeman-fit}
          \label{Code_Fit_I}

\begin{figure}[t]
     \begin{center}
     \includegraphics[scale=0.1]{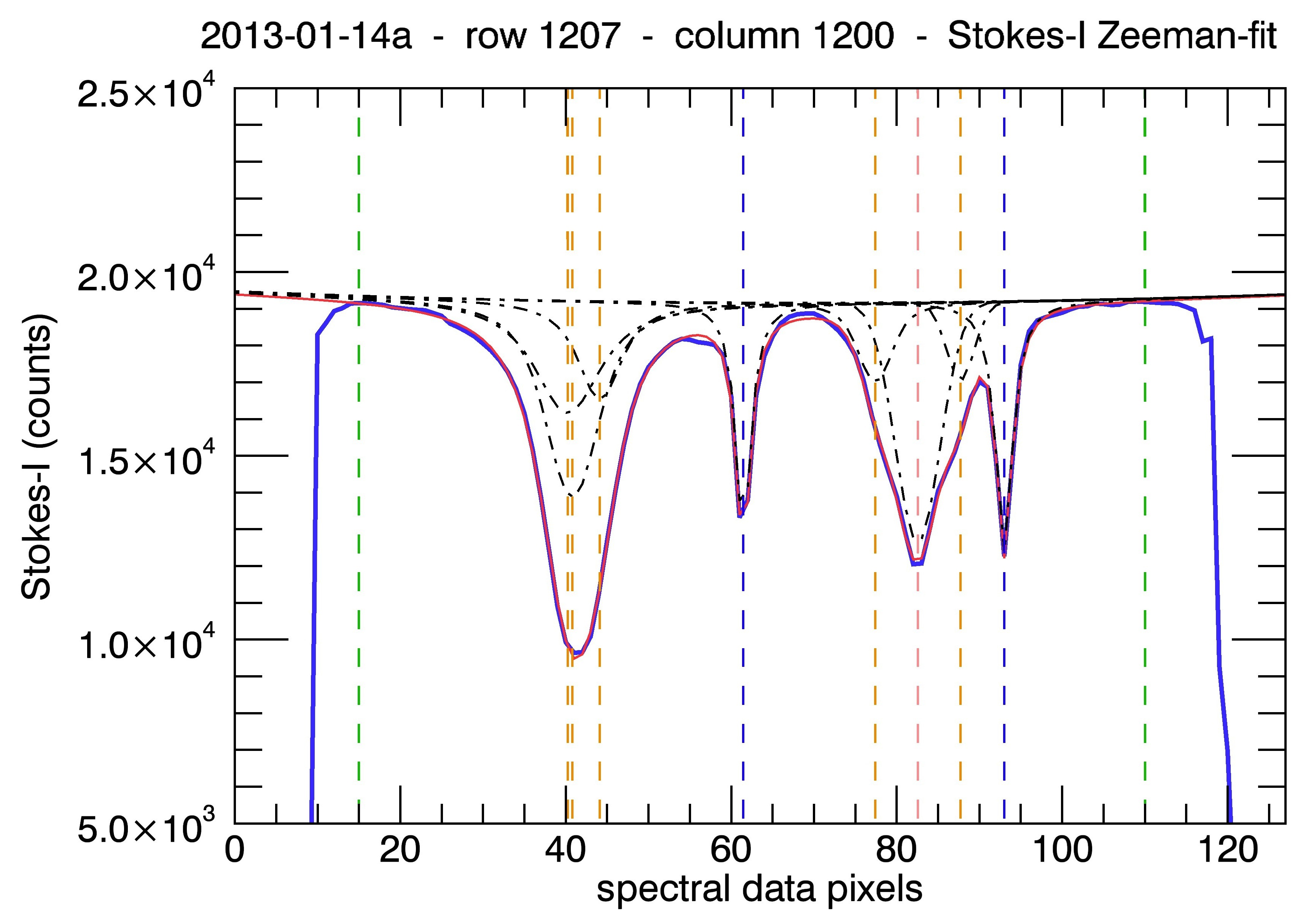}
     \caption[Stokes-I Zeemanfit profile example]{Data profile (blue) and fit (red) of a high-field-strength Stokes-I spectrum fit with 5 Voigts ($x_0$: orange dashes) for all of the 6302\AA \ Fe-triplet lines except for the 1 gaussian ($x_0$: pink dashes) fit to the 6302.5\AA-$\pi$-component, and 2 Lorentzians ($x_0$: blue dashes) for the telluric O$_2$ lines.  The black dash-dot lines plot the fit components.  The green dashes denote the range-bounds of the fit data; note that this example fit was run while the code was in development, when the data range was hardcoded to [15:110].}
     \label{FIG_IFitZeeman}
     \end{center}
     \end{figure}

An example of the fit produced by the default settings of the Zeeman-fit to the Stokes-I profile (SOLIS\_VSM\_ZEEMANFIT\_IFIT\_6302V() ) is shown in Figure~\ref{FIG_IFitZeeman}.
The fit includes:
\begin{list}{*}{}
     \item a polynomial-fit for the background of order 2
     \item 2 Lorentzian profiles defining the telluric O$_2$ lines
     \item 1 Gaussian profile defining the $\pi$-component of the 6302.5\AA \ Zeeman-triplet
     \item 5 Voigt profiles used to define both the two $\sigma$-components of the 6302.5\AA \ Zeeman-triplet and the three lines allotted to fitting the 6301.5\AA \ Zeeman-broadened line-set.
\end{list}
Beyond this profile definition, the fit conforms to the following constraints:
\begin{list}{*}{}
     \item The separation parameters between the $\pi$ and $\sigma$-component line-center positions (the $\Delta x_\sigma$ parameters) (for the 6302.5\AA \ line) are restricted to positive values only, defined such that this constraint maintains the wavelength ordering: $x_{0,\sigma_V} < x_{0,\pi} < x_{0,\sigma_R}$.
     \item The amplitudes of the 6302.5\AA \ $\sigma$-components are fit as some fraction, $c_{A,R/V}$, each of the $\pi$-component amplitude, and each $c_A$-fit-value is constrained to the limits $0.0 \leq c_A \leq 1.0$ so that neither $\sigma$-component amplitude will be fit larger than the $\pi$-component amplitude.  This amplitude restriction is used when reasonable, and is lifted in any cases where the initial guess for the $\pi$-amplitude is less than either or both of the guessed $\sigma$-amplitudes.
     \item All fit line-amplitudes are required to be less than zero (absorption lines).
\end{list}

The initial input for the continuum-polynomial coefficients comes from a straight-line fit to the highest points in each half of the input data-profile.  The initial guesses for the line amplitudes are measured from this straight line down to the data-value at the guessed line-center positions.

The initial-guesses for the line-center positions of the two telluric O$_2$ lines and the two Fe $\pi$-component lines are determined first by a call to solis\_vsm\_ZEEMANFIT\_FLOODMIN(), which locates profile data ranges for four data-minimum 'pools' and returns the minimum-value pixel location and range-center location for each pool.  The data-minimum positions are used for the two O$_2$ lines, and the pool-center position is used for the 6301.5 $\pi$-component line.  For the 6302.5 $\pi$-component position, the code first redefines the 'data-pool' boundary to be at a level that is halfway between the local data-minimum value and the lowest of the two local-data-peak values between the Fe-triplet set and the telluric O$_2$ lines on either side.  This level tends to be well centered around the lines composing this Fe-triplet set, and it is the center of {\em this} pool that is used to guess the position of the 6302.5 $\pi$-component line.

\begin{figure}
     \begin{center}
     \includegraphics[scale=0.26]{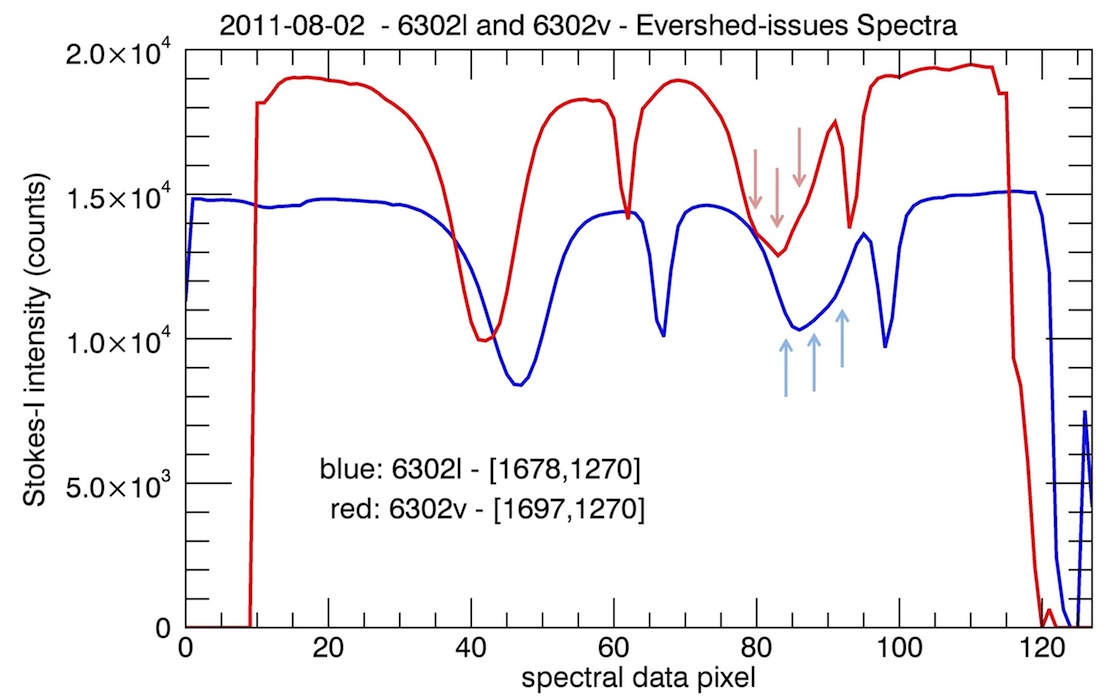}
     \caption[6302l versus 6302v spectra - Evershed affected]{Two Stokes-I spectra taken from the same umbral region of a near-limb sunspot, one from a SOLIS 6302l observation and the other from a 6302v observation taken roughly three hours later.  Between the two instruments, spectral features appear offset relative to each other on the camera.  The arrows indicate the approximate locations of the three Zeeman-components of the 6302.5 triplet set for each of the spectra.  Note that for this particular example, the $\pi$-component for the 6302l profile appears to be suppressed relative to that seen in the 6302v profile.  The sunspot region sampled is the area most clearly effected by issues relating to the Evershed effect (see \S\ref{Code_TripletPos}).}
     \label{FIG_EvershedSpectra}
     \end{center}
     \end{figure}

In earlier versions of the code, the data-minimum positions were used to guess the positions for all four of the above lines.  However, this was changed when the code was modified to accommodate the processing of 6302l spectra, for which, as shown in Figure~\ref{FIG_EvershedSpectra},
certain regions/spectra have $\pi$-components of insufficient strength the make the early method always viable.

Next, determining appropriate guesses for the four $\sigma$-component line-center positions is much trickier, and our selection of the most appropriate method is the subject of \S\ref{Code_TripletPos}.  The default method in the pipeline is to use the Level-2-corrected continuum-intensity value.  This value is compared to the Level-2 fit for the quiet-sun limb-darkened  continuum profile to make an observational-studies-based estimate of the local magnetic-field-strength and therefore of the appropriate Zeeman-triplet line-separation.  (See section \S\ref{Code_Calc} for details on the appropriate line-separation--to--Field-strength conversion equations.)

\begin{figure}
     \begin{center}
     \includegraphics[scale=0.3]{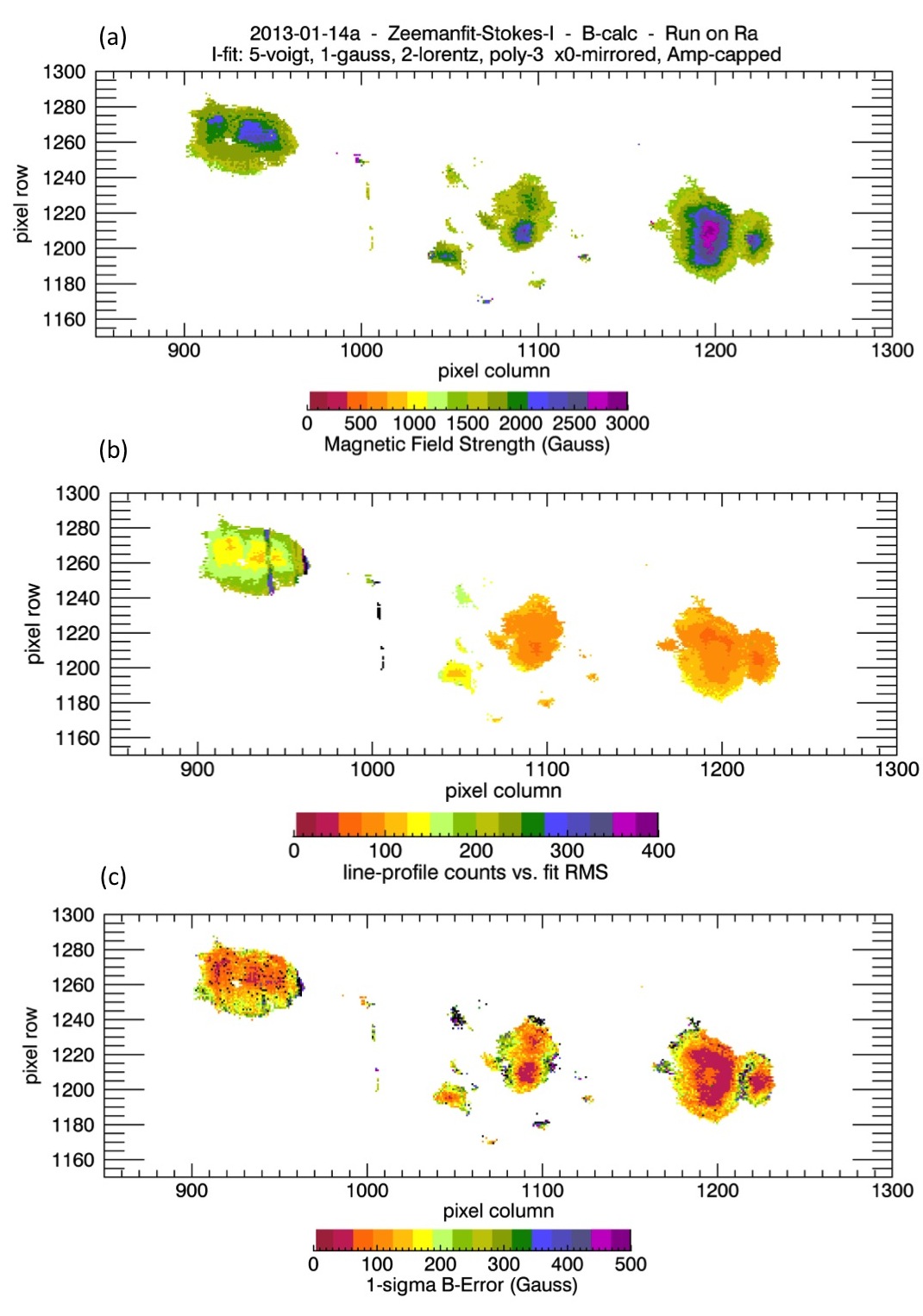}
     \caption[Stokes-I-calculations maps]{Maps generated using the fit to the Stokes-I profile of the magnetic-field-strength (a), the RMS of the fit-line to the spectral-data profile (b), and the 1-sigma error calculated due to the uncertainty in the fit line separations (c).  (Note that these calculations were run during testing with a polynomial background of order 3.)}
     \label{FIG_imageStokesImap}
     \end{center}
     \end{figure}

The guesses for the Gaussian-widths-values ($\sigma$) for the triplet lines are set at a fraction of the line-center separations.  For all lines that use it, the guesses for the Lorentzian-width-values ($\gamma$) are set to 1.0 pixel.  
Basic results of using this model can be seen in Figure~\ref{FIG_imageStokesImap}.

{\bf Now, in terms of why the profile definitions specified above were chosen:}
\begin{list}{*}{}
     \item The {\bf background polynomial of order 2} was chosen as being likely sufficient to account for the background continuum and unlikely to introduce ugly artifacts into the line fits.  However, the time has not been taken to determine whether a lower-order background might be equally sufficient or even superior.
     \end{list}
     \begin{list}{*}{}
     \item {\bf Lorentzians} were chosen to fit the the {\bf telluric O$_2$ lines} for simplicity, reduction of parameter space, and because they appear to do a perfectly lovely job of it.
    \item All of the lines used to fit the Stokes-I Zeemanfit profiles are required to be {\bf in absorption} because all of the features expected to appear in these spectra are absorption features.  However, particularly at low continuum intensities, the blue half of the spectra are sometimes fairly bumpy, leading to occasional in-emission fits when positive line amplitudes were previously allowed (PROVER2B = 1.1).
     \end{list}
     \begin{list}{*}{}
     \item An early-development version of the code required that the positions of the two 6302.5 $\sigma$-components be mirrored about the $\pi$-component (that $\Delta x_{\sigma_V} = \Delta x_{\sigma_R}$).  Requiring this mirroring allowed for a small reduction in parameter space that is physically reasonable given the Zeeman-splitting mechanism and was found to produce less noisy results.  However, as is discussed in detail in \S\ref{Code_TripletPos}, this mirroring was discarded in order to avoid certain artifacts occurring near the solar limb.
     \item The constraint that says 6302.5\AA \ triplets that show the capacity for it should be fit {\bf requiring the $\pi$-component amplitude to be larger} than either of the 
     $\sigma$-component amplitudes is a slippery one.  If the observations contained no scattered light, the amplitude of the $\pi$-component would depend on the amount of both unpolarized and linearly polarized light, and could be small.  However, testing suggested that including this constraint helps to produce less noisy $B$-strength maps, and apparently more robust fit computations.  Furthermore, a significant amount of scattered light {\bf is} expected to be present in the data (particularly in the dark umbral centers where the field alignment is most liable to present longitudinal).  Please compare the results from Figure~\ref{FIG_imageStokesImap} with those in Figure~\ref{FIG_imageNotAmpCapped},
     \begin{figure}[t]
     \begin{center}
     \includegraphics[scale=0.15]{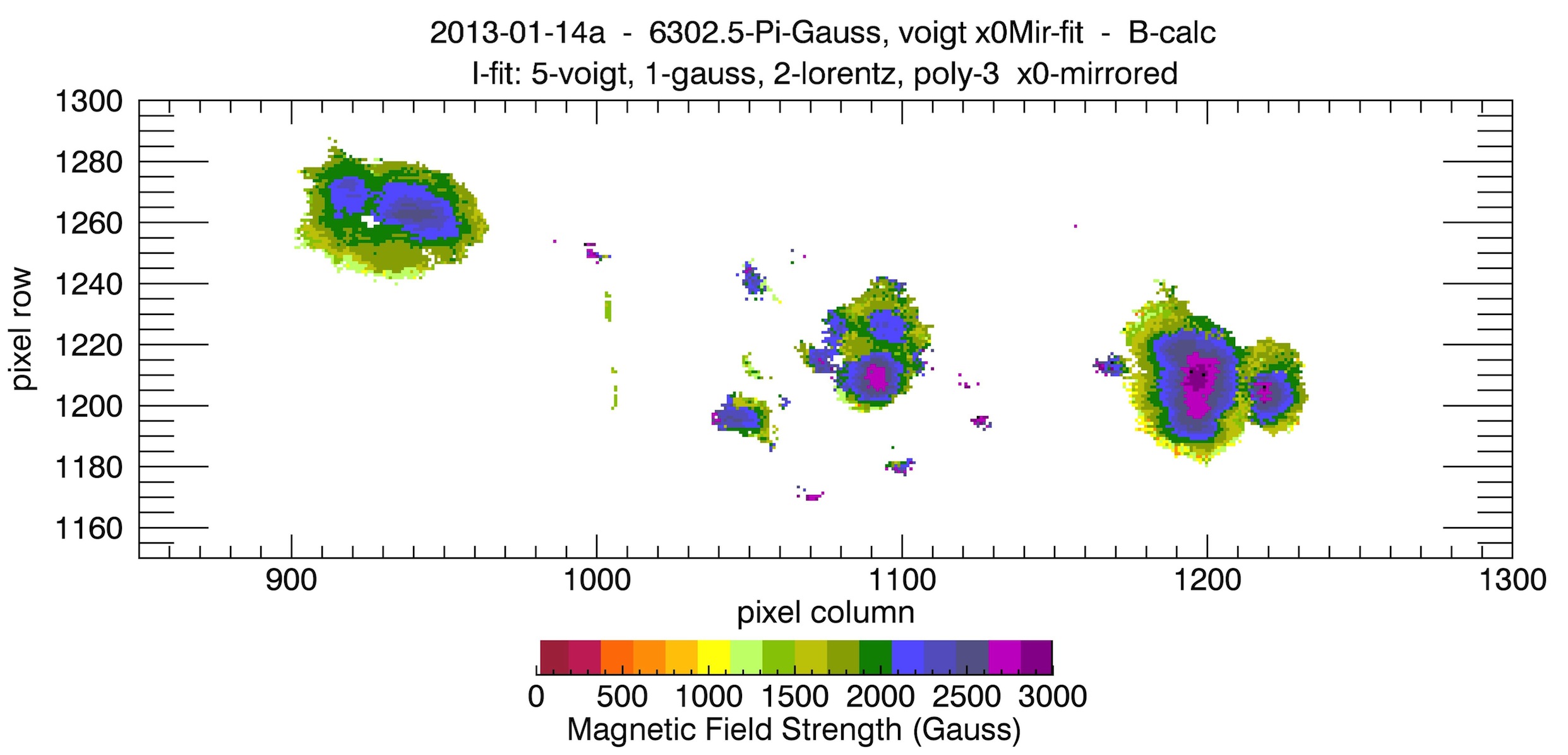}
     \caption[B-calculation map - amplitudes not restricted]{A map of computed magnetic-field strength using a fit to the Stokes-I profile that does {\bf not} include restricting the 6302.5\AA \ $\sigma$-component amplitudes to be lower than the $\pi$-component amplitude.  Note the large $B$-values computed for many of the spatially outlying pixels.}
     \label{FIG_imageNotAmpCapped}
     \end{center}
     \end{figure}
which presents a $B$-strength map generated using early fits (using /Vguess, see \S\ref{Code_TripletPos}) run without this amplitude-restricting constraint.
     \end{list}
     \begin{list}{*}{}
     \item Finally, the single most important constraint for acquiring usable Stokes-I $B$-calculations is that the 6302.5\AA \ {\bf $\pi$-component line be fit as a Gaussian only}, not as a Voigt.  Figure~\ref{FIG_imageStokesI6gauss}
     \begin{figure}
     \begin{center}
     \includegraphics[scale=0.3]{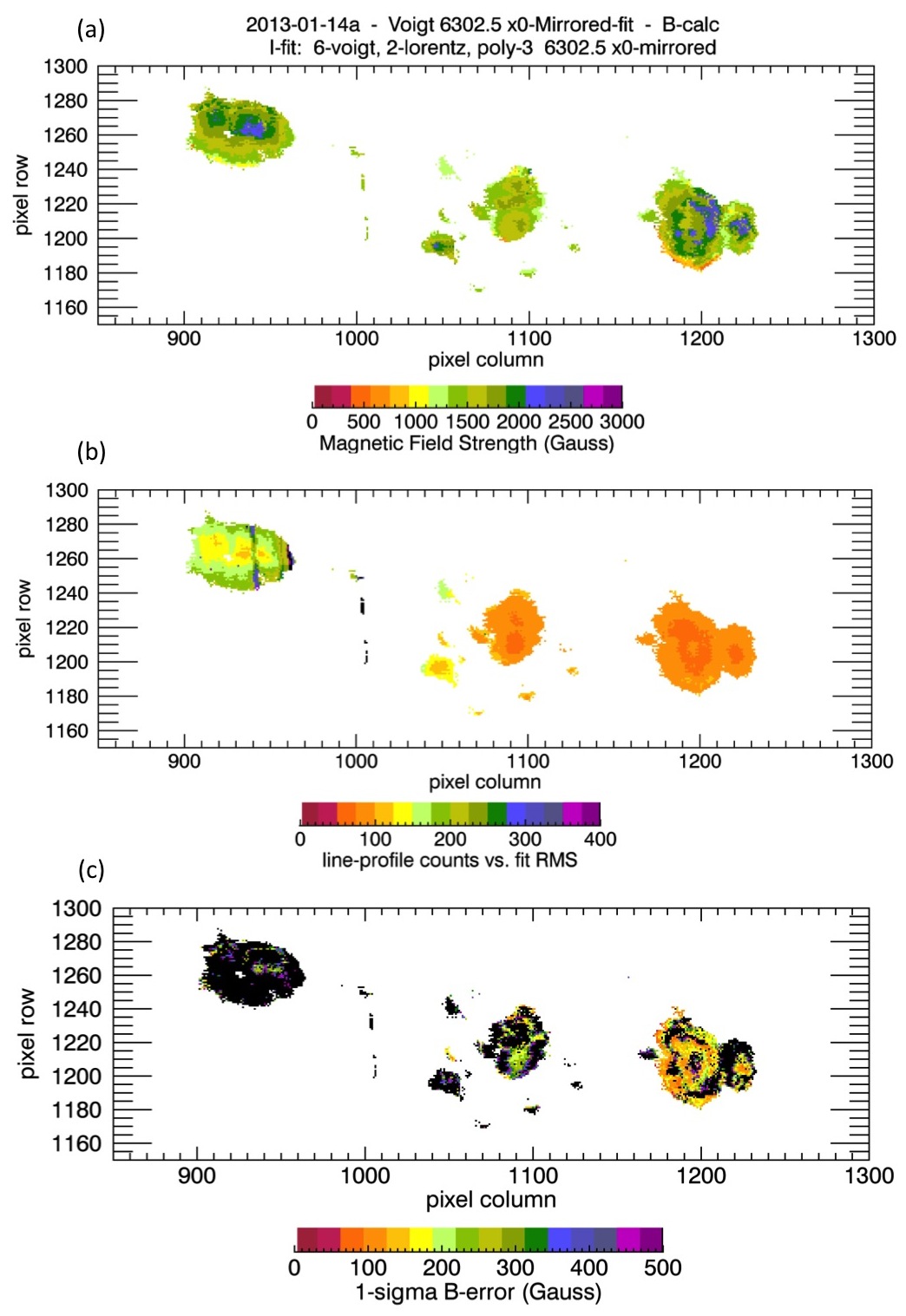}
     \caption[Stokes-I-calculations maps for fit with 6-Voigts]{Maps generated using 6 Voigts to fit the the Fe-triplet lines in the Stokes-I profile (rather an using 5 Voigts plus 1 Gaussian for the 6302.5\AA~$\pi$-component): the calculated magnetic-field strength (a), the RMS of the fit-line to the spectral-data profile (b), and the 1-sigma error calculated due to the uncertainty in the fit line separations (c).}
     \label{FIG_imageStokesI6gauss}
     \end{center}
     \end{figure}
presents the results of running the Stokes-I fit using all Voigts for the solar triplet lines.  While the rms of the fit itself can appear relatively low and lovely with this method (panel b), the uncertainty on the $\sigma$-component line positions is actually very high, resulting in a large uncertainty in the $B$-calculation (panel c).  Most notably, though, the structure of the magnetic-field strength across the dark umbral regions is implausible, with reported umbral field strengths often dipping across sun-spot centers, rather than peaking (panel a).  Compare that result to that of Figure~\ref{FIG_imageStokesImap}, where the $\pi$-component-as-a-Gaussian constraint has been instituted.  The results toward the penumbral regions are noisier than might be preferred (but this method of fitting to Stokes-I is chosen for being most robust in the strong-field cases), but the field-strength--profile morphologies agree with the morphology suggested by the continuum image (Figure~\ref{FIG_Iactive}),
     \begin{figure}
     \begin{center}
     \includegraphics[scale=0.14]{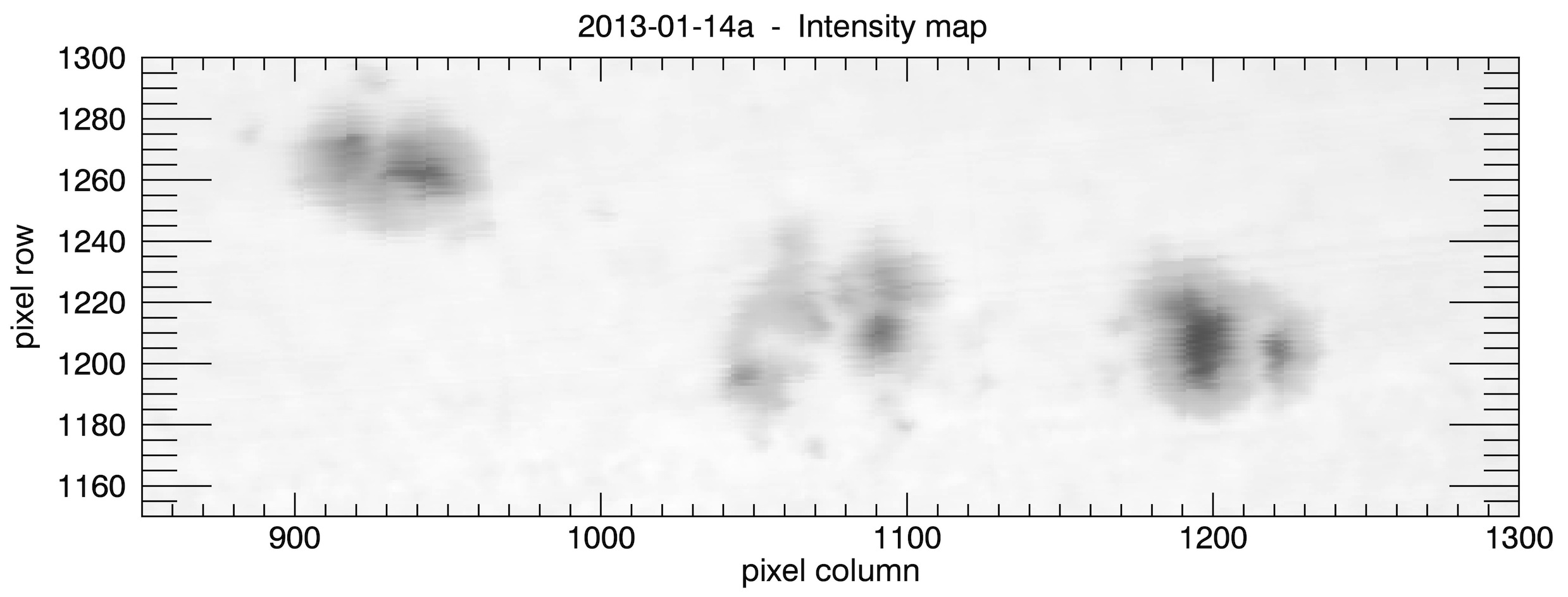}
     \caption[Active Region Intensity Map]{Map of the continuum intensity across the active region modeled in the preceding figures.}
     \label{FIG_Iactive}
     \end{center}
     \end{figure}
and the uncertainty in the fit line-centers (and therefore the $B$-calculation) has dropped into reasonable bounds.
\end{list}

          \subsubsection[Stokes-V fit]{Stokes-V Zeeman-fit}
          \label{Code_Fit_V}

\begin{figure}
     \begin{center}
     \includegraphics[scale=0.17]{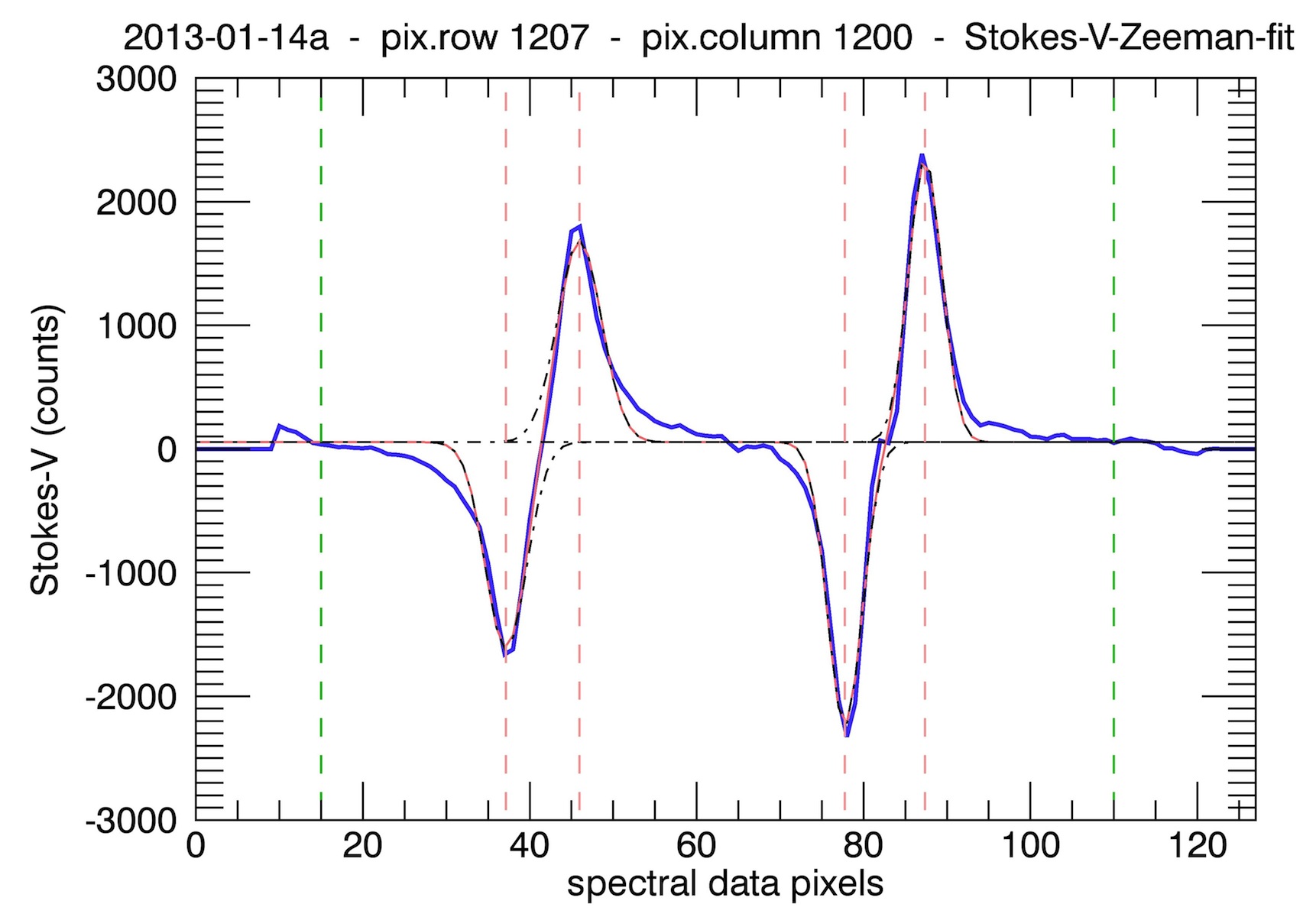}
     \caption[Stokes-V Zeemanfit profile example]{Data profile (blue) and fit (red) of a high-field-strength Stokes-V spectrum fit with 4 gaussians ($x_0$: pink dashes) for the 6302\AA \ polarized $\sigma$-component lines.  The black dash-dot lines plot the fit components.  The green dashes denote the range-bounds of the fit data.}
     \label{FIG_VFitZeeman}
     \end{center}
     \end{figure}

\begin{figure}[t]
     \begin{center}
     \includegraphics[scale=0.35]{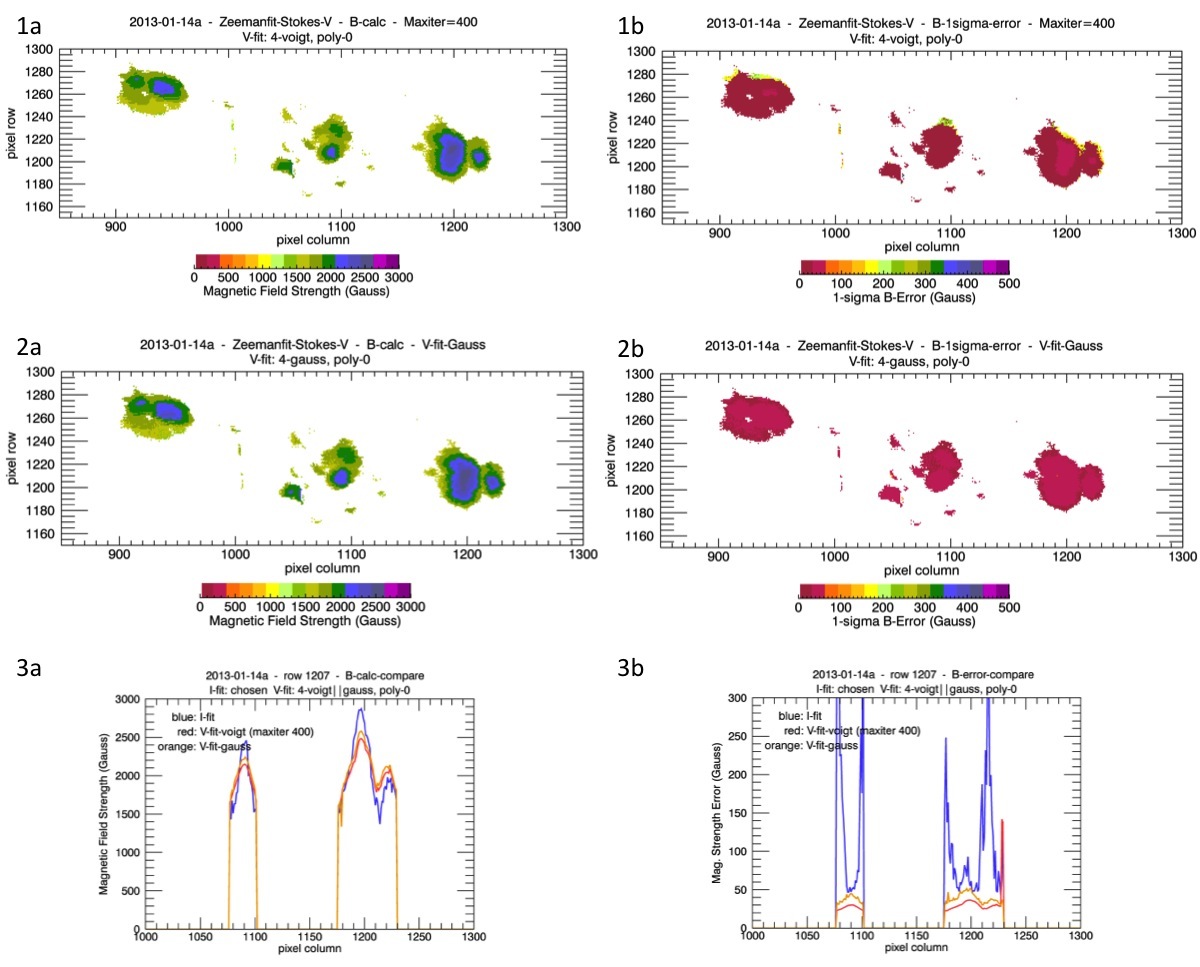}
     \caption[Stokes-V Zeemanfit maps: Voigt-fit vs. Gauss-fit]{Mapped comparison of calculated B (column a) and B error (column b) using fits to the Stokes-V profiles.  Row 1 depicts mapped results using fits with Voigt profiles, Row 2 depicts mapped results using (pipeline-adopted) Gaussian profiles, and Row 3 presents the Strength and Error profiles for one pixel row across a sunspot, comparing the results from the Stokes-I-fit (blue), the Stokes-V-Voigt-fit (red) and the Stokes-V-Gauss-fit (orange).}
     \label{FIG_VFit_GaussVVoigt}
     \end{center}
     \end{figure}

An example of the fit produced by the default settings of the Zeeman-fit to the Stokes-V profile (SOLIS\_VSM\_ZEEMANFIT\_VFIT\_6302V() ) is shown in Figure~\ref{FIG_VFitZeeman}.
The fit includes:
\begin{list}{*}{}
     \item a constant background (polynomial order 0)
     \item 4 Gaussian profiles
\end{list}
Beyond this profile definition, the fit is entirely unconstrained, and the results of this model can be seen in Figure~\ref{FIG_VFit_GaussVVoigt}.

The initial guesses for the positions and amplitudes of the four fit lines are taken from the min/max positions and values in each half of the data profile, and the guess for the background constant is entered as 0.  As with the Stokes-I Zeeman-fit, the guesses for the Gaussian widths ($\sigma$) are set to a fraction of the separations between each of the min/max peak pairs.

Though Gaussians do not fit the Stokes-V line profiles as well as Voigts do, the choice was made to fit using only Gaussians for the sake of both speed and robustness.  Running the code using Voigts tended, for several hundred pixels across the map, to max out the number of iterations run by MPFITFUN() (using either the default MAXITER=200, or MAXITER=400).  Switching to Gaussians not only reduces the number of fit parameters, it also drastically cuts down the number of pixels that run maximum iterations (to only a handful across the whole map, default MAXITER).

Also, running the fit with Voigts produces noticeably larger uncertainty in the line-center positions around some of the penumbral edges, where the Stokes-V counts are not as strong.  See Figure~\ref{FIG_VFit_GaussVVoigt}
for a comparison of map results produced by Stokes-V fits using Voigts versus Gaussians.  One note: running the Stokes-V fit with Gaussians does tend to produce roughly 4\% larger Stokes-V $B$-values than does running the fit with Voigts, whatever that ultimately means.

          \subsubsection[Stokes-I Non-zeeman-fit]{Stokes-I Non-zeeman-fit}
          \label{Code_Fit_out}

\begin{figure}[t]
     \begin{center}
     \includegraphics[scale=0.22]{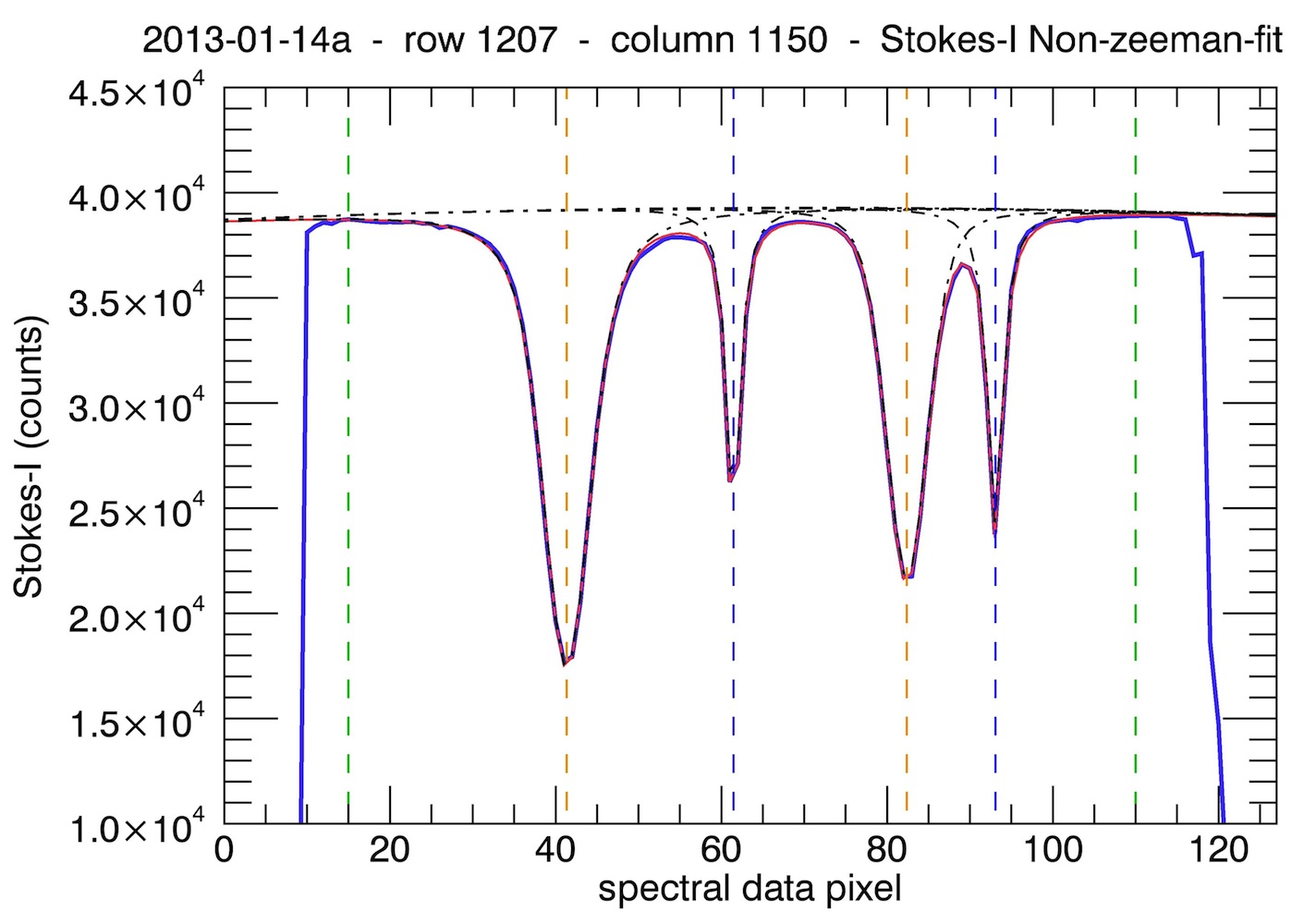}
     \caption[Stokes-I Non-zeeman-fit profile example]{Data profile (blue) and fit (red) of a low-field-strength Stokes-I spectrum fit with 2 voigts ($x_0$: orange dashes) for the 6302\AA \ Fe lines and 2 lorentzians ($x_0$: blue dashes) for the telluric O$_2$ lines.  The black dash-dot lines plot the fit components.  The green dashes denote the range-bounds of the fit data.}
     \label{FIG_IFitNonzeeman}
     \end{center}
     \end{figure}

An example of the fit produced by the default settings of the Non-zeeman-fit to the Stokes-I profile (SOLIS\_VSM\_ZEEMANFIT\_OUTFIT\_6302V() ) is shown in Figure~\ref{FIG_IFitNonzeeman}.
The fit includes:
\begin{list}{*}{}
     \item a polynomial-fit for the background of order 2
     \item 2 Lorentzian profiles defining the telluric O$_2$ lines
     \item 2 Voigt profiles defining the solar Fe lines (6301.5\AA \ \& 6302.5\AA)
\end{list}
Beyond this profile definition, the fit is entirely unconstrained.

As during the Zeeman Stokes-I fit (\S\ref{Code_Fit_I}), the initial input for the polynomial coefficients comes from a straight-line fit to the two highest points in each half of the data-profile; and the guessed line amplitudes are measured from this straight line down to the data-value at the guessed line-center positions.

For this fit, all line-center positions are acquired via a call to \_ZEEMANFIT\_FLOODMIN() in order to locate the four primary profile minima within the chosen data-fit range.

As to why the profile definitions specified above were chosen:
\begin{list}{*}{}
     \item As with the Stokes-I Zeeman-fit, the {\bf polynomial order 2} was chosen as being likely sufficient to account for the background continuum and unlikely to introduce ugly artifacts into the line fits.  However, the time has not been taken to determine whether a lower-order background might be equally sufficient or even superior.
     \item {\bf Lorentzians} were chosen to fit the the {\bf telluric O$_2$ lines} for simplicity, reduction of parameter space, and because they appear to generally do a perfectly lovely job of it.
     \item {\bf Voigts} were chosen to fit the {\bf solar Fe lines} because they could be.  For the vast majority of cases, Gaussians would probably do a perfectly acceptable job, considering that the positions of the telluric lines are the only results of interest when running this fit.  However, it is possible that the results would be less robust if this fit code, using Gaussians for the solar lines, were run on an image pixel that {\bf ought} to be Zeeman-triplet-fit but wasn't due to any issues with the pixel-selection criteria (see \S\ref{Code_Sorting}).
\end{list}

This Non-zeeman fit is used only to calculate the dispersion for those pixels not fit as high-field-strength Zeeman-triplet profiles (and only if the /DISPERCALC keyword is flagged).  For further details concerning the dispersion choices, calculations, and results, see \S\ref{Code_Disper}.

     \subsection[\textcolor{blue}{Triplet-lines Separation: Initial Guess and Fit Constraints}]{\textcolor{blue}{Triplet-lines Separation: Initial Guess and Fit Constraints}}
     \label{Code_TripletPos}

The separation of the triplet-model $\sigma$-component lines is the parameter of greatest importance to the Zeemanfit results (see \S\ref{Code_Calc}) and must be handled with a great deal of care in order to ensure that both the code and the results remain:
\begin{enumerate}
\item consistent
\item robust
\item reasonable when compared to continuum sunspot morphology
\end{enumerate}
This is fairly straight-forward when running a fit on the Stokes-V profile.  As long as the polarized signal is strong enough to be both well above the noise and minimally affected by more complex structure, the $\sigma$-component line-center positions  should be resolved to match fairly closely with the four min/max profile peaks in each half of the Stokes-V profile.

However, in the Stokes-I profile, the $\sigma$-component line-centers do not have profile minimums with which to directly correspond, and, so far, four separate techniques have been tested for providing a guess as to the Stokes-I $\sigma$-component positions to solis\_vsm\_ZEEMANFIT\_MULTVOIGTFIT() and MPFITFUN().  These techniques and our concluding choices are discussed below, along with the pros and cons of requiring the fit 6302.5 $\sigma$-component positions to be mirrored about the $\pi$-component.  Ultimately, we have chosen a default operation that uses the relative continuum intensity to make a guess of the magnetic-field strength and therefore of the $\sigma$-component line separation, and to {\em not} require that the fit $\sigma$-component positions be mirrored about the fit $\pi$-component position.

\begin{figure}
     \begin{center}
     \includegraphics[scale=0.2]{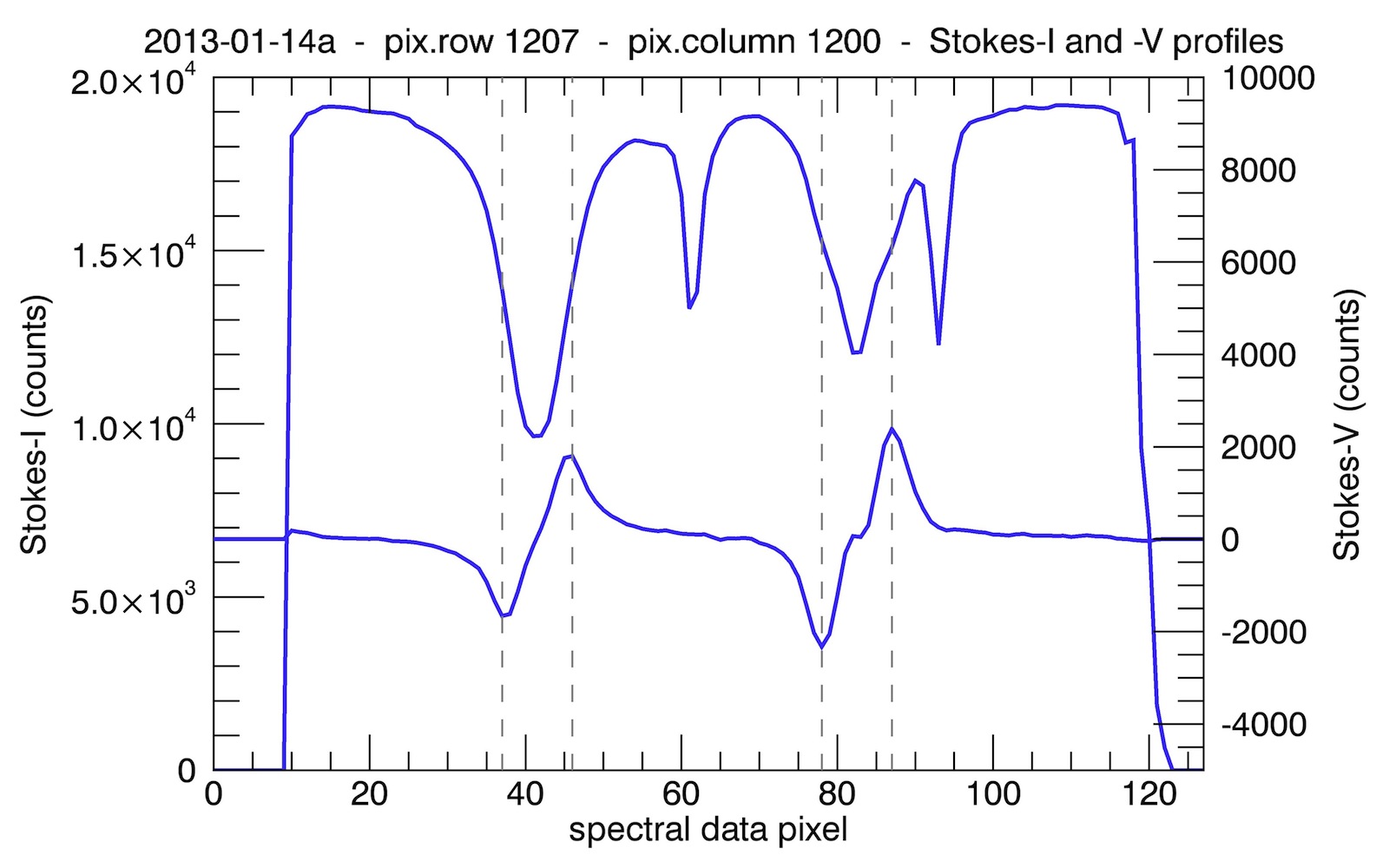}
     \caption[Stokes-I $x_0$ guesses from Stokes-V data positions]{Stokes-I and -V profiles denoting the use of the Stokes-V min/max locations (grey dashes) for guessing the positions of the Stokes-I--Fe-triplet--$\sigma$-component line centers if the keyword /Vguess is flagged or PFILE has not been specified.}
     \label{FIG_Vminmax}
     \end{center}
     \end{figure}

Because it is both readily available and affected directly by properties of the local magnetic field, the first technique tried for guessing the Stokes-I $\sigma$-component positions was to use the same input positions as for the Stokes-V fit, the four min/max peaks of each half of the V-profile, as depicted in Figure~\ref{FIG_Vminmax}.
Non-standard operation of the code using /Vguess does still allow for this use of the Stokes-V profiles in guessing the Stokes-I triplet-lines center positions, and is the necessary mode of operation if the code is computing fits for Level-0.5 spectra that do not have corresponding Level-2-corrected continuum-intensity values readily available with which to guess the separation.  However, there are two primary drawbacks to using this /Vguess method.
\begin{enumerate}
\item Even in regions of strong total magnetic field, the Stokes-V signal is not always strong enough for this guess method to be employed.  Particularly near the solar limb, strongly transverse fields may mean a weak or non-existent polarized signal, disallowing results to be computed for many pixels across, particularly, the dark umbral regions of a sunspot.
\item Even where the polarized signal is perfectly clear and strong, it does not directly correlate to the {\em total} magnetic-field strength being sought in a fit to the Stokes-I profile.  The Stokes-V peak separation depends on one component of the magnetic-field, and using it to guess the Stokes-I $\sigma$-component separation leads to random inconsistencies in how the Stokes-I-fit calculations proceed.  As outlined below, the Stokes-I results are sensitive to the initially guessed separation.
\end{enumerate}
However, using the /Vguess method does appear to be reasonably robust.  It will be biased toward reporting underestimates of the total magnetic-field strength.

\begin{figure}
     \begin{center}
     \includegraphics[scale=0.125]{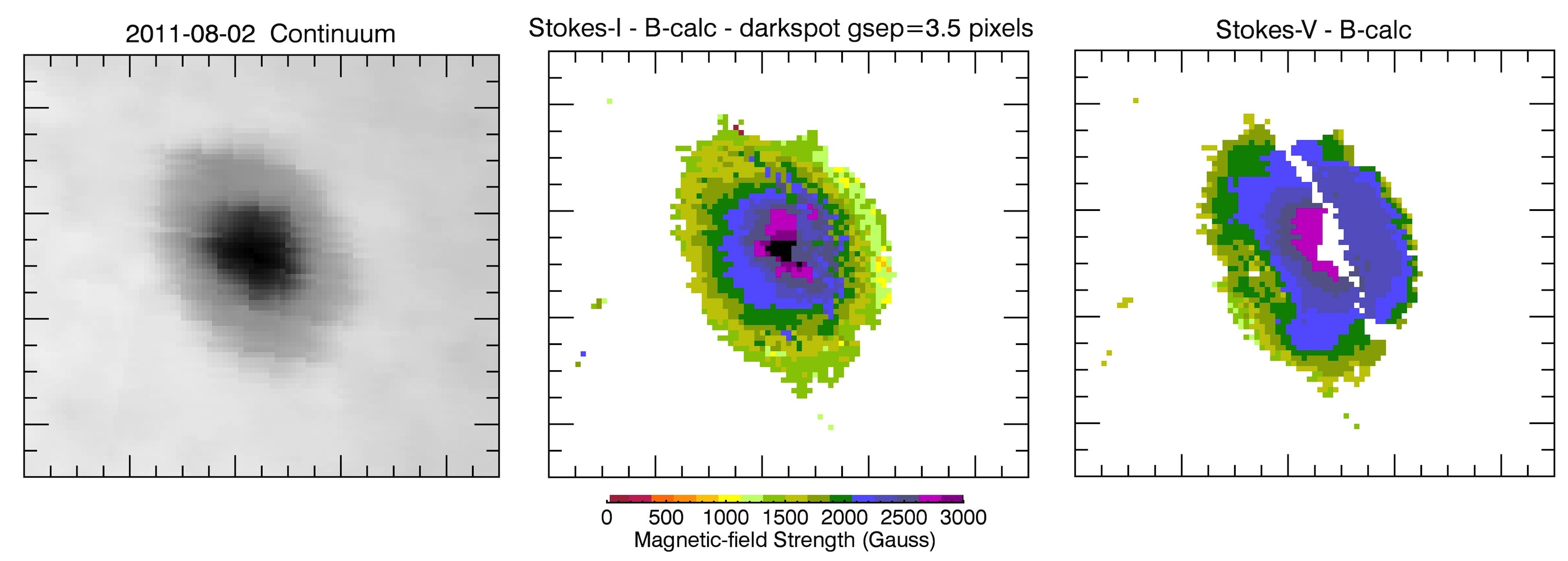}
     \caption[$B$-values returned across a near-limb sunspot]{Calculated magnetic-field-strength values returned from fits to the Stokes-I (center panel) and Stokes-V (right panel) spectra for a near-limb sunspot.  Stokes-V results show a gap where Stokes-V counts fall below the Stokes-V selection criteria.   This gap coincides with the position of the magnetic neutral line in the sunspot, where the field is perpendicular to the line-of-sight.  The Stokes-I results are filled in across the gap using a guessed $\sigma$-$\pi$ separation of 3.5 pixels, producing a clear discontinuity in umbral results.}
     \label{FIG_darkspotdiscontinuity}
     \end{center}
     \end{figure}

Next, a simplistic method is to merely make a fixed-width guess for the $\sigma$-component line separation applied to every Stokes-I profile where necessary.  This was initially done to try to fill in missing regions of Stokes-I results left blank by the Stokes-V-guess method, and an example of this can be seen in Figure~\ref{FIG_darkspotdiscontinuity}.
A $\sigma$-$\pi$-guessed separation of 4.0 pixels corresponds to a guessed total separation of 8.0 pixels, or a 6302.5\AA-triplet-computed field-strength of $\sim2080$ Gauss.  

If we had no other recourse for guessing the $\sigma$-component line positions, guessing this 4-pixel separation for every computation (and {\em not} using any other guess method for some pixels) would be a viable method.  It meets the consistency requirement easily, and at this separation the code running on the VSM-6302-vector data is reasonably robust (though note that this has only been tested while 6302.5 $\sigma$-component mirroring was required.)  It has problems with the outer penumbral regions, but these regions are of significantly less interest than the umbral centers.  

\begin{figure}
     \begin{center}
     \includegraphics[scale=0.38]{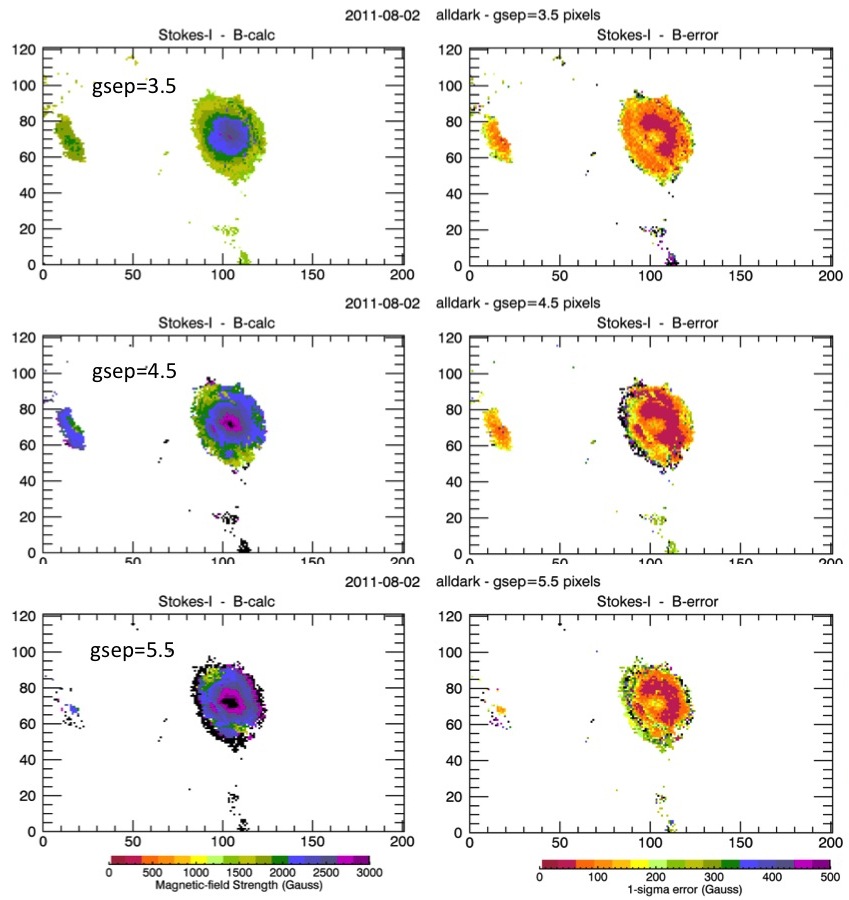}
     \caption[Dark-pixel varied position-guessing results]{The calculated magnetic-field strengths (left panels) and errors (right panels) for Zeemanfit-Stokes-I calculations run using a fixed $\pi$-$\sigma$-separation guess (increasing from the top panels to the bottom) for each pixel calculation in a given run.}
     \label{FIG_BcalcGsep}
     \end{center}
     \end{figure}

However, as is clearly depicted in Figure~\ref{FIG_BcalcGsep},
the results of the Zeemanfit-computations are quite sensitive to the initially guessed separation (to the initial guess of the magnetic-field strength).  If the initial guess is high, the outputted result tends to be high, and if the initial guess is substantially higher than physical reality, the computation becomes unstable.  We could keep our guess everywhere very modest, but then results computed for umbral centers are often well below their physical values.  We need a guess method that is not constant, but is directly and consistently tied to the observations.

Therefore, two other guess methods were tested.  One used an algorithm called solis\_vsm\_ZEEMANFIT\_FWHM() to input the profile-minimum location returned from solis\_vsm\_ZEEMANFIT\_FLOODMIN() corresponding to the guessed $\pi$-component position and return a point to either side corresponding to the half-maximum (minimum really since these are absorption features) positions of that region of the spectral profile.  The other method takes in the background-relative continuum intensity value for that image pixel and uses that value to estimate the local magnetic-field strength and therefore the $\sigma$-$\pi$ line-center separation for the Stokes-I profile, following the equation suggested by Jack Harvey:
\begin{equation}
   B_\mathrm{estimate}
=
   1500 + 1500 \left( 1 - \left(I/I_{LD}\right) \right)
\, ,
\label{EQ_Iguess}
\end{equation}
where $I_{LD}$ is the limb-darkened, background intensity computed in equation~\ref{EQ_limbdark}.  (See \S\ref{Code_Calc} for the equations converting between line-separation and magnetic-field strength.  The dispersion used in making this estimate is always the dispersion-value assigned to the camera (Rockwell or Sarnoff) on which the data were taken.)

The \_FWHM() method provides a very consistent treatment of the data and requires nothing outside of the Stokes-I profile itself to proceed.  Unfortunately, it is not well suited to providing $\sigma$-position guesses in the case that the $\sigma$-component-absorption is on the same order or stronger than the $\pi$-component absorption, and it is subject to provide over-estimates in the cases when the line-set is strongly influenced by broadening mechanisms outside the Zeeman effect.  

One such case appears to be the effect of spatial averaging of fine structure from the Evershed effect, noticed primarily in the computed results for strong sunspots near the solar limb (though exactly which sunspots and how close to the limb appears to be fairly variable, and possibly dependent on data quality).  False-high-field artifacts probably related to this effect (they always appear on the sun-center-ward side of a sunspot) seem to be linked primarily to an asymmetry that appears in the observed positioning of the $\sigma$-component line-centers relative to the $\pi$-component center.  Therefore, the effect is exacerbated by requiring that the profile-fit mirror the positions of the $\sigma$-components about the $\pi$-component line center.  (Remember, the $\sigma$-component {\em amplitudes} remain unconstrained relative to each other, and therefore the fit might return results where one component amplitude is quite small compared to the other.)

\begin{figure}
     \begin{center}
     \includegraphics[scale=0.225]{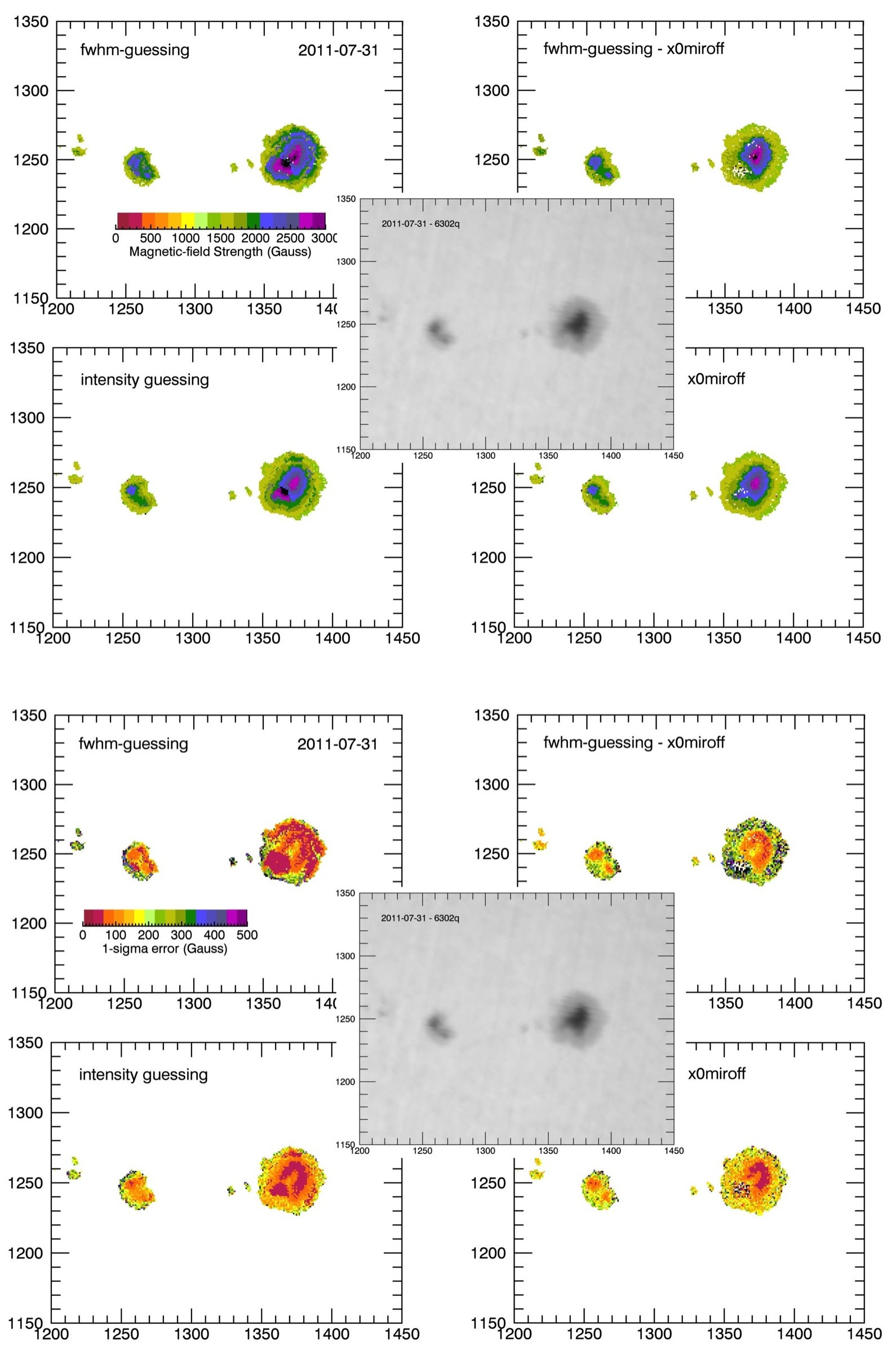}
     \caption[Comparison of four fit configurations for an Evershed-affected sunspot - Bmag]{{\bf Top set:} The computed magnetic-field strengths for four different configurations of the Stokes-I--fit modeling upon an Evershed-artifact affected sunspot.  Top panels model the fit using the half-maximum profile-positions to guess the $\sigma$-component line-center positions.  Bottom panels guess the separation based on the Level-2 corrected image-pixel continuum intensity (shown in the center panel).  Left panels require 6302.5 $\sigma$-positions be mirrored about the $\pi$-center position.  Right panels were run with this mirroring requirement turned off.  {\bf Bottom set:} The corresponding computed 1-sigma errors.}
     \label{FIG_EvershedCompMag}
     \end{center}
     \end{figure}

Computed Stokes-I magnetic-field-strength values and 1-sigma errors are presented in Figure~\ref{FIG_EvershedCompMag}
for four fit-computation scenarios, i.e., varying both the guess-method used to estimate the $\sigma$-component line-center positions and whether or not the fit requires the 6302.5 $\sigma$-positions to be mirrored about the $\pi$-center position.  From these plots (mirroring: left versus no-mirroring: right), the advantages and disadvantages of mirroring become clear.  Requiring mirrored positions removes one (of many) parameters from the list of those that need to be fit, and the results tend to be more stable.  Further, for observations primarily affected by the the Zeeman effect, such a condition is reasonably physical.  {\em However}, the artifact related to the Evershed effect is clearly present under the mirroring condition in those sunspots so influenced, and furthermore, the {\em error} reported in the regions of this artifact  is quite low, leading one to suspect that under these conditions the mirroring requirement over-constrains the fit, leading to a narrow best-fit space that is never-the-less unrealistic.

\begin{figure}
     \begin{center}
     \includegraphics[scale=0.23]{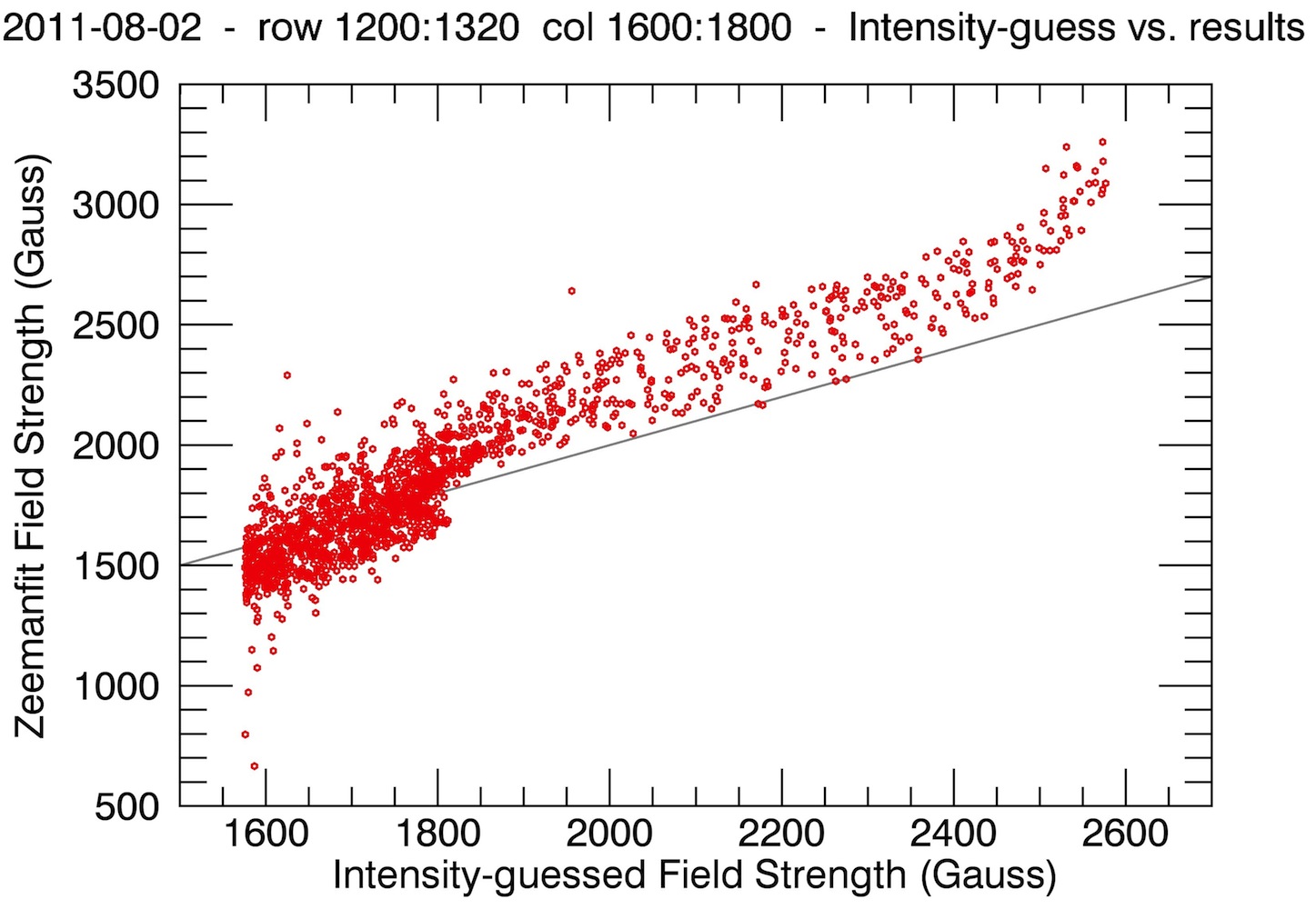}
     \caption[Scatter plot - intensity guess versus fit result]{Scatter plot of the Zeemanfit-computed Stokes-I results for one sunspot versus the input field-strength guesses based on the continuum intensity.  The grey line represents the location of 1:1 correspondence.}
     \label{FIG_Iscatter}
     \end{center}
     \end{figure}

However, the higher degree of instability that results when the mirroring condition is turned off is a non-negligible problem, as can be seen by the number of missing pixel results in the pipeline-selected default fit in the lower-righthand panels.  This fit is more stable than the half-maximum-guessing counterpart (top-right), but these pixels have still been clipped out of the results due to very high 1-sigma error ($>$ 1000 Gauss) and/or excessively high computed--field-strength values ($>$ 5000 Gauss).  In fact, the intensity estimate for the magnetic-field strength (equation~\ref{EQ_Iguess}) usually underestimates the total strength, as can be seen in the Figure~\ref{FIG_Iscatter}
 scatter plot of computed strength versus guessed strength.  Increasing the estimate of the scattered-light contribution in the observations would make a better match between input-guess and output results.  However, running Zeemanfit on several days of observations with a higher guess estimate resulted in a higher degree of instability, which, in the case of one sunspot, severely reduced the number of pixel-results reported in the umbral center.  Therefore, the default fit parameters we have chosen to operate under represent a compromise necessitated by using a parameter-intensive, brute-force technique on of-order-90-pixels worth of spectral data per profile.

     \subsection[\textcolor{blue}{Dispersion}]{\textcolor{blue}{Dispersion}}
     \label{Code_Disper}

The wavelength-per-pixel dispersion, used in the calculation of the magnetic-field-strength, can come from either of two sources, detailed below.  It can be specified by the code according to the camera the data was taken on (\S \ref{Code_Disper_Camera}), or it can be computed separately at each pixel location via the separation of the two telluric O$_2$ lines (\S \ref{Code_Disper_Calc}).  Default operation of the code assumes that the camera dispersion is known and is supplied by the code according to the camera used to gather the ingested data.

          \subsubsection[\textcolor{blue}{Camera-defined Dispersion}]{\textcolor{blue}{Camera-defined Dispersion}}
          \label{Code_Disper_Camera}

Currently for the 6302v and 6302l data, the code uses the dispersion values supplied by Lorraine Callahan:
\begin{list}{*}{}
\item 0.02409 \AA/pixel for SARNOFF data
\item 0.0271 \AA/pixel for ROCKWELL data.
\end{list}

\begin{figure}[t]
     \begin{center}
     \includegraphics[scale=0.4]{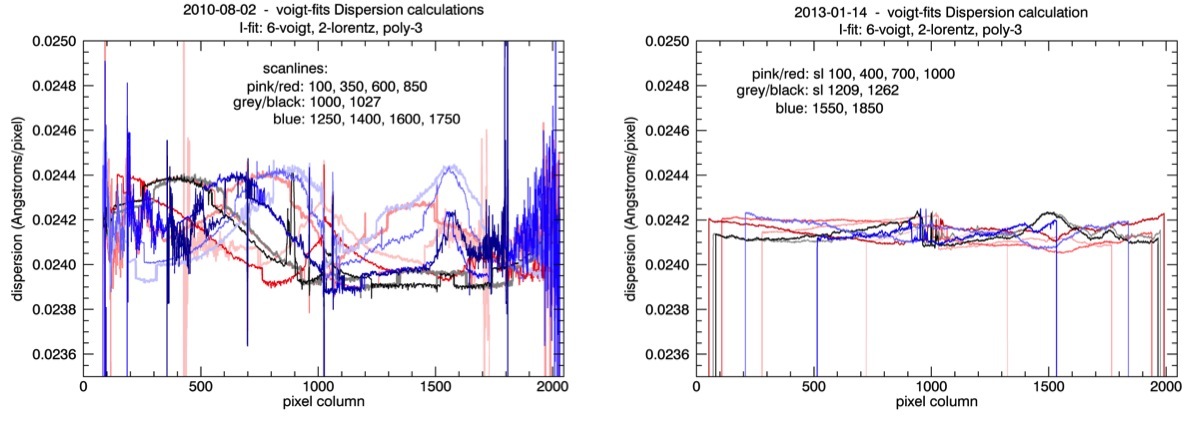}
     \caption[Comparison of Calculated Dispersion on Different Days]{Comparision of dispersion-calculation results across several latitudes for two different days of 6302v data.  Note that these calculations were made before the final Stokes-I--Zeeman-fit setup was decided upon, but that the positions of the telluric O$_2$ lines are quite stable across variations in the fit setup.}
     \label{FIG_comparedispercalc}
     \end{center}
     \end{figure}

For an error value on the SARNOFF data, the code supplies an error estimate of 1.3\% that we made by inspection of the Stokes-I-calculated dispersion (see \S~\ref{Code_Disper_Calc}) returned for a number of scan-lines of 6302v data taken on Aug.~2, 2010.  In Figure~\ref{FIG_comparedispercalc},
compare the dispersion calculated for that day to the dispersion calculated for several scan-lines of Jan.~13, 2013 data.  The August 2010 data reports a noticeably larger spread in the calculated dispersion, and we've chosen 1.3\% relative error as a likely upper-limit.  Note that the dispersion value is probably better well known than this, but this is the error-value the code is using right now.

No ROCKWELL data has yet been processed by this code, therefore the estimated error on the dispersion for ROCKWELL data is currently set to 0.74\%.  This is taken purely from Lorraine's observation that some of the reprocessing code uses a hard-wired, "default" dispersion-value of 0.0271 \AA/pixel, while the database, used mostly by older code, lists 0.0273 \AA/pixel for the ROCKWELL camera.

          \subsubsection[\textcolor{blue}{Calculated Dispersion per-Image-Pixel}]{\textcolor{blue}{Calculated Dispersion per-Image-Pixel}}
          \label{Code_Disper_Calc}

     \begin{figure}
     \begin{center}
     \includegraphics[scale=0.18]{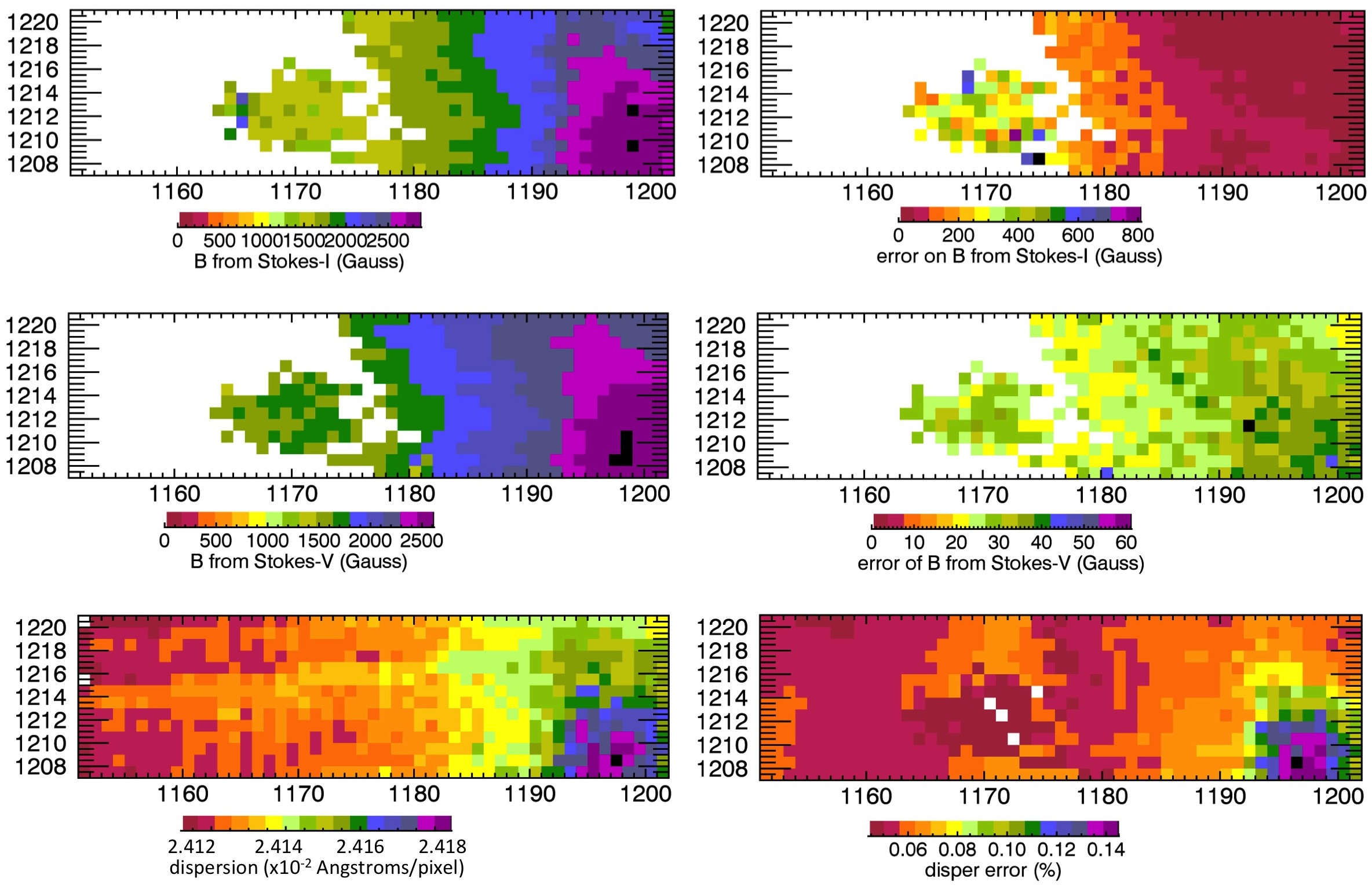}
     \caption[Map of all Zeemanfit output]{The bottom row maps a sample of the dispersion calculated using the Zeemanfit and \_OUTFIT\_6302V() code.  The left panel shows the value of the dispersion, which varies smoothly across the frame (between 0.02412 and 0.02418 \AA/pixel).  The right panel shows the \%-error on the dispersion calculation, which shows a patch of lower-error (white and dark-red pixels) in the shape of the B-calculated pixels (above panels)  The dispersion for {\em those} pixels is calculated following the Stokes-I fit by \_IFIT\_6302V().}
     \label{FIG_dispermap}
     \end{center}
     \end{figure}
     \begin{figure}
     \begin{center}
     \includegraphics[scale=0.2]{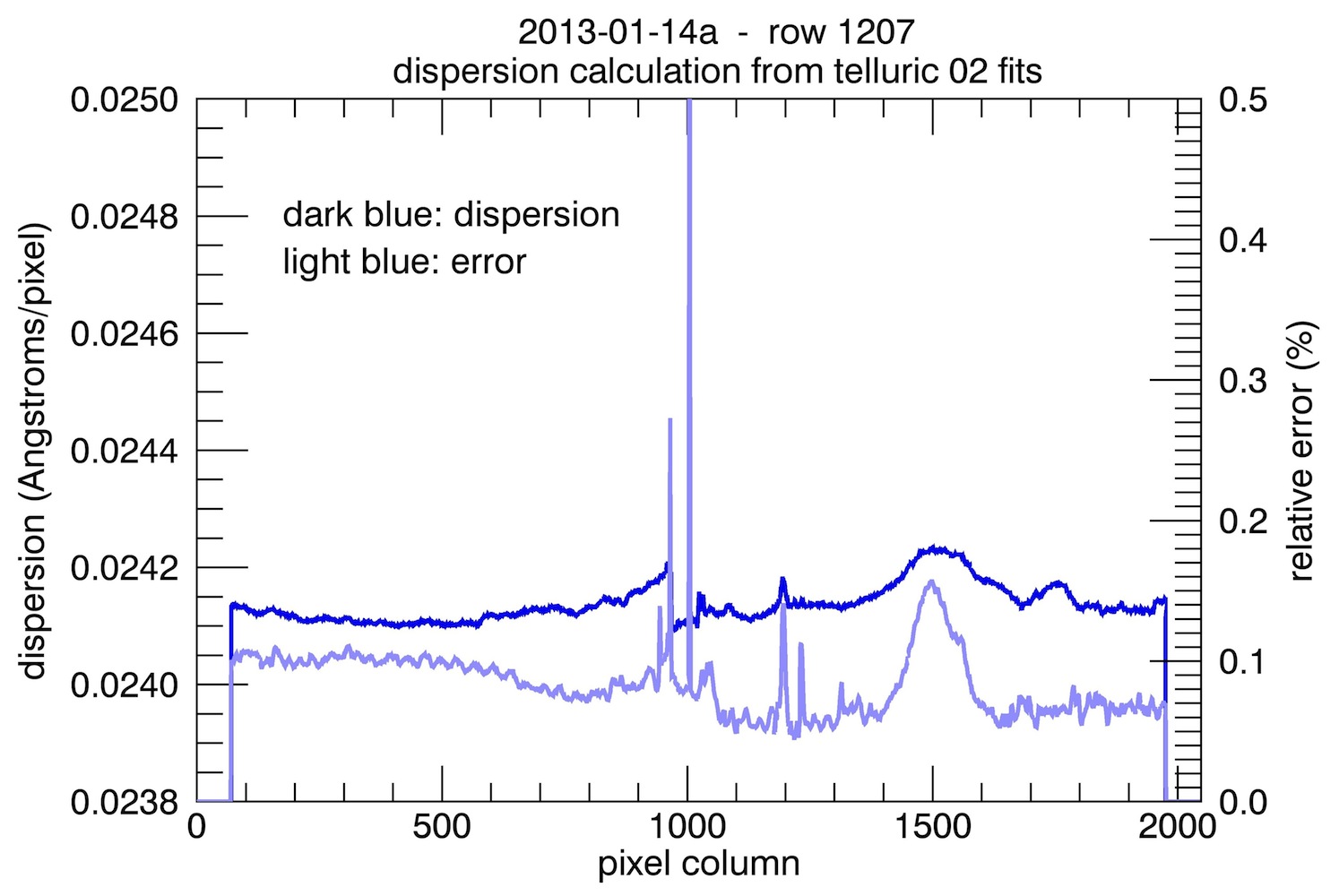}
     \caption[Calculated Dispersion and Error]{The dispersion (dark blue) calculated by the code across a single scan line, plus the corresponding relative error (light blue).}
     \label{FIG_1disperror}
     \end{center}
     \end{figure}

When running the code, one may choose to set the Keyword flag /DISPERCALC.  If so, the magnetic field-strength is calculated using the dispersion value calculated from the Zeeman-fit to the Stokes-I profile (solis\_vsm\_zeemanfit\_IFIT\_6302V() ), {\bf plus} all of the non-Zeeman-selected pixels also have a (simplified) fit run on their Stokes-I profiles (solis\_vsm\_zeemanfit\_OUTFIT\_6302V() ) so that the dispersion values and errors may be returned for every solar-disk--image pixel.  One {\bf important note}, then, is that the calculated dispersion values presented across a given scan line or image are coming (randomly, as far as the distribution of active regions is random) from calculations using two different Stokes-I fitting schemes (see Figure~\ref{FIG_dispermap}).

The dispersion is assumed to be constant across the spectral frame.  It is calculated using the positions of the two in-frame telluric O$_2$ lines as:
\begin{equation}
   d
=
   \frac{\lambda_{O_{2b}} - \lambda_{O_{2a}}}
          {x_{0,O_{2b}} - x_{0,O_{2a}}}
\, ,
\end{equation}
where $d$ is the dispersion in \AA/pixel, $x_{0,O_{2b}}$ and $x_{0,O_{2a}}$ are the line-center positions for the telluric lines returned from the fit to the Stokes-I profile, and for this data range $\lambda_{O_{2b}}=6302.7629\AA$ and $\lambda_{O_{2a}}=6302.0005\AA$ according to the Pierce-Breckenridge Atlas (ftp://nsokp.nso.edu/pub/atlas/linelist/).

The 1-sigma relative error on the above-calculated dispersion is calculated as:
\begin{equation}
   \frac{\sigma_d}
           {d}
=
   \frac{ \sqrt{ \sigma_{xO_{2b}}^2 + \sigma_{xO_{2a}}^2 } }
          {\left| x_{0,O_{2b}} - x_{0,O_{2a}} \right|}
\, ,
\end{equation}
where $\sigma_d$, $\sigma_{xO_{2b}}$, and $\sigma_{xO_{2a}}$ are the 1-sigma uncertainties on the dispersion, and the O$_{2b}$ and O$_{2a}$ line-center positions, respectively.  See \S \ref{Code_Calc} for comments regarding the acquisition of the $\sigma_{x0}$-values from MPFITFUN(), though this is not as convoluted for the telluric O$_2$ line positions as it is for the Zeeman-fit Stokes-I 6302.5\AA-triplet positions.

An example of the dispersion calculated across a scan line is presented in Figure~\ref{FIG_1disperror}.
One important note is that the O$_2$ lines are not well resolved by this data.  Therefore, the dispersion value calculated according to their fit separations is likely noisier than any real dispersion variation.

     \subsection[Magnetic-Strength Calculation]{Magnetic Strength and Error Calculations}
     \label{Code_Calc}

Once the line profiles have been fit, and the pixel dispersion either chosen or calculated, the magnetic-field-strength can be calculated (for regular-Zeeman-split lines) from the formula:
\begin{equation}
   B
=
   \frac{\left| \lambda_{0,\pi} - \lambda_{0,\sigma} \right|}
           {4.67\times10^{-5} \,\, g \,\, \lambda_{0,\pi}^2}
\, ,
\end{equation}
where, $B$ is the magnetic-field strength (in Gauss), $g$ is the Lande factor appropriate to the particular Zeeman-triplet, $\lambda_{0,\pi}$ is the center wavelength of the principal line (in cm), and $\lambda_{0,\sigma}$ is the center wavelength of one of the Zeeman-triplet $\sigma$-component lines (in cm).

Because we assume a constant dispersion value across the full (128-pixel) spectral image, and because we may or may not be using a fit-symmetry about the $\lambda_{0,\pi}$ component, the implementation in the code becomes:
\begin{equation}
   B
=
   \frac{10^{-8}}
           {4.67\times10^{-5} \,\, g \,\, \lambda_{0,\pi}^2}
   \, * \, d
   \, * \, \frac{1}{2} \left( x_{0,\sigma_R} - x_{0,\sigma_V} \right)
\, ,
\label{EQ_Bcalc}
\end{equation}
where $d$ is the dispersion in \AA /pixel, $\lambda_{0,\pi}$ is the nominal wavelength in cm of the line of interest, $x_{0,\sigma_R}$ and $x_{0,\sigma_V}$ are the respective fitted pixel positions of the right and left $\sigma$-components, and $g$ is the Lande factor (2.5 in the case of the 6302.5\AA \ Fe line).

If the above equation is applied to the $\sigma$-component-line positions derived from the fit to the Stokes-I profile, then the computed $B$-value should represent the {\bf total} magnetic-field strength.  However, if the equation is applied to the line-peak positions derived from the fit to the Stokes-V profile, then $B$ should be a calculation of the {\bf longitudinal} magnetic-field-strength only.

Note that the code {\bf does} include a magnetic-field-strength calculation for the 6301.5\AA \ line ($g = 1.67$).  However, it should mostly just be ignored.  This is because:
\begin{enumerate}
     \item The 6301.5\AA \ line is Zeeman-anomalous; physically, it's not as simple as a triplet.
     \item The fitting code does not impose any constraints that would make the 6301.5\AA \ "triplet" lines behave at all like a Zeeman triplet.  They are allowed to vary nearly entirely unconstrained to fit the data profile in whatever way seems best.  Not infrequently, they wind up in a different wavelength order than the one originally specified, i.e., the $\pi$-component can wind up situated outside rather than between the two $\sigma$-components.
\end{enumerate}
For these reasons, the fit to the Fe 6301.5\AA\, is considered unphysical.

\begin{figure}[t]
     \begin{center}
     \includegraphics[scale=0.23]{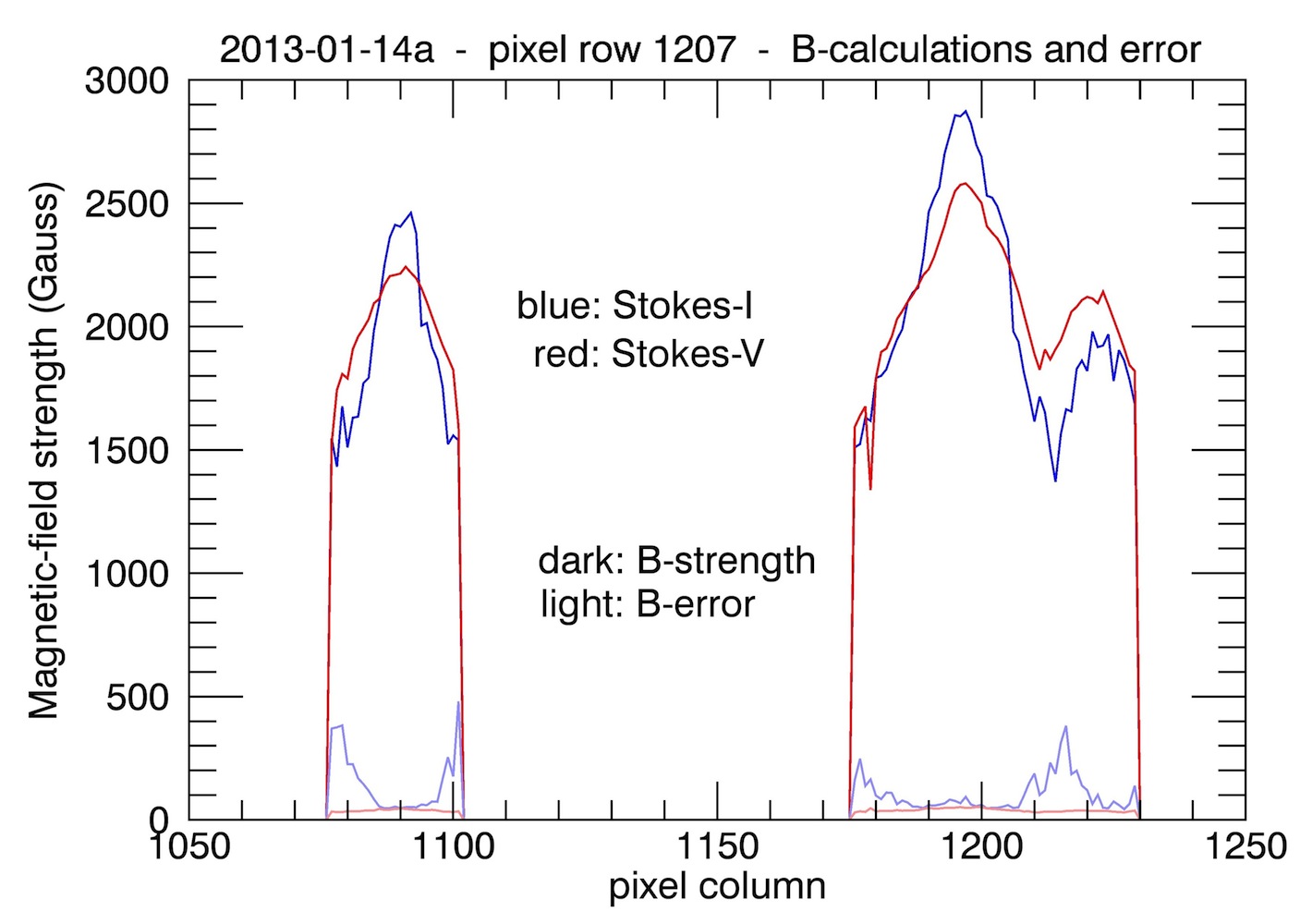}
     \caption[Example B-calc's and errors]{The magnetic-field strengths and uncertainties for an example test run (using /Vguess, \S\ref{Code_TripletPos}) calculated from the Stokes-I and -V spectra across an active region.}
     \label{FIG_Bcalcanderror}
     \end{center}
     \end{figure}

Figure~\ref{FIG_Bcalcanderror}
presents an example of calculated Stokes-I and -V $B$-values and errors for one pixel row across an active region.  Note that uncertainty in the Stokes-I result increases toward the edges of the sunspots, in keeping with this Zeeman-fit method, which is aimed at fitting results in the high-magnetic-field-strength regions where the splitting is nice and broad.

The 1-sigma error calculated for the reported $B$-values takes into account the uncertainty on both the dispersion and the fit line-center positions.  In general, it is given by:
\begin{equation}
   \sigma_B^2
=
   \left(B \, \sigma_d\right)^2
\,+\, 
   \left(c_B d \, \sigma_{\Delta x_{0,R}}\right)^2
\,+\,
   \left(c_B d \, \sigma_{\Delta x_{0,V}}\right)^2
\, ,
\end{equation}
where $\sigma_B$, $\sigma_d$, $\sigma_{\Delta x_{0,R}}$, and $\sigma_{\Delta x_{0,V}}$ are the 1-sigma error estimates on the magnetic-field strength, dispersion, and right- and left-$\sigma$-component separations from $x_{0,\pi}$, respectively; in units of Gauss, \AA/pixel, and pixels, respectively.  $c_B$ is the coefficient of constants given in equation~\ref{EQ_Bcalc}: $\frac{1}{2}\times 10^{-8} / \left( 4.67\times 10^{-5} \, g \, \lambda_{0,\pi}^2 \right)$.

The above equation is used to calculate error for the Stokes-V-calculated magnetic-field strengths, as well as for the Stokes-I strengths in the default processing where the $x_0$-mirroring has been turned off so that all of the line-center positions are allowed to freely vary.  However, optional operation of the code (using the SMIRROR keyword) {\em does} allow the Stokes-I 6302.5 positions to be mirrored (the separations from $x_{0,\pi}$ and the two $x_{0,\sigma}$'s are forced to be equivalent).  Therefore, the equation of the error in that instance becomes:
\begin{equation}
   \sigma_B^2
=
   \left(B \, \sigma_d\right)^2
\,+\, 
   4 \left(c_B d \, \sigma_{\Delta x_{0,\sigma}}\right)^2
\, ,
\end{equation}

Because the data-profile values passed into MPFITFUN() do not have particular error values associated with each data-point, MPFITFUN() does not return a reliable 1-sigma estimate of the error on the fit parameters.  However, according to the documentation in MPFITFUN(), {\em "If} you can assume that the true reduced chi-squared value is unity [...]~then the estimated parameter uncertainties can be computed by scaling PERROR by the measured chi-squared value":
     \begin{verbatim}
                   PCERROR = PERROR * SQRT(BESTNORM / DOF)
     \end{verbatim}
Therefore, the code retrieves the PERROR-vector and the BESTNORM and DOF values from MPFITFUN() and uses the above equation to compute the $\sigma_{x,0}$ values returned from running the various line fits.  

Note also that, while \_ZEEMANFIT\_MULTGAUSSFIT() does return a parameter array containing the $x_{0,\sigma_R}$ and $x_{0,\sigma_V}$ parameters, for the Zeeman-fit to the Stokes-I profiles, MPFITFUN() itself varies ${\bf \Delta x_{\sigma_R}}=x_{0,\sigma_R} - x_{0,\pi}$, and ${\bf \Delta x_{\sigma_V}}=x_{0,\pi}-x_{0,\sigma_V}$ in the specific cases of the two 6302.5\AA \ $\sigma$-component line-centers.  Therefore, the 1-sigma values passed back for those two parameters correspond to these later quantities in the Stokes-I calculations.

\section[\textcolor{blue}{Functional Comparison to Outside Results}]{\textcolor{blue}{Functional Comparison to Outside Results}}
\label{Compare}

Comparison of field strengths derived by this code with observations from Bill Livingston (1.5 micron near-infrared), Hinode (high spatial resolution in Fe I 6302.5A), and HMI/SDO (Fe I 6173.3~\AA) indicate a good agreement with this method.  A detailed description of this comparison will be published in a separate paper.

\section[\textcolor{blue}{Future Work}]{\textcolor{blue}{Future Work}}
\label{Future}

\begin{list}{*}{}
     \item {\bf ROCKWELL processing:}  The Zeemanfit code has not been tested on any Rockwell-era data.  There are at least a couple of issues to consider/fix if at some point the code will be applied to Rockwell data or to any future data acquired on a different camera.
        \begin{list}{*}{}
        \item {\bf Spectral-data pixel values:} The \_FITRANGE() algorithm uses a number of hard-coded values to account for things like the size of the input-spectrum data vector and very roughly where certain features are expected to lie on that data profile.  For processing using a different camera, those pixel values may need to be tailored to the camera or some other method selected for specifying those ranges.  Also, in solis\_vsm\_zeemanfit\_IFIT\_6302V.pro, pixel-index 70 is used as a midway mark for a backup call to \_FLOODMIN(); again, this may need to be modified or per-camera tailored.
        \item {\bf Dispersion: }  The camera-set dispersion fixed into the code for data taken with the ROCKWELL camera is 0.0271\AA/pixel, the value currently being used in the reprocessing code.  However, Lorraine Callahan has noted that the value in the database is 0.0273\AA/pixel.  Therefore, fairly arbitrarily, the relative-error on this dispersion value that we have coded in is this 0.74\% difference.  However, we have not had a chance to run this code in dispersion-calculation mode (or any other) on any ROCKWELL data (if there are any ROCKWELL spectra around to try it on), so these choices remain pretty random at the moment and should be verified before any Zeemanfit 'reprocessing' of ROCKWELL data is implemented.
        \end{list}
     \item {\bf Last bits and bobs:}  There are still a couple of issues of not-quite-satisfaction with the code-results such as:
     	\begin{enumerate}
	\item Sometimes the solver might choose {\bf $\sigma$-amplitude = 0} for one $\sigma$-component, etc.  See Figure~\ref{FIG_wierdlinefits}
	\begin{figure}[t]
     	\begin{center}
     	\includegraphics[width=\textwidth]{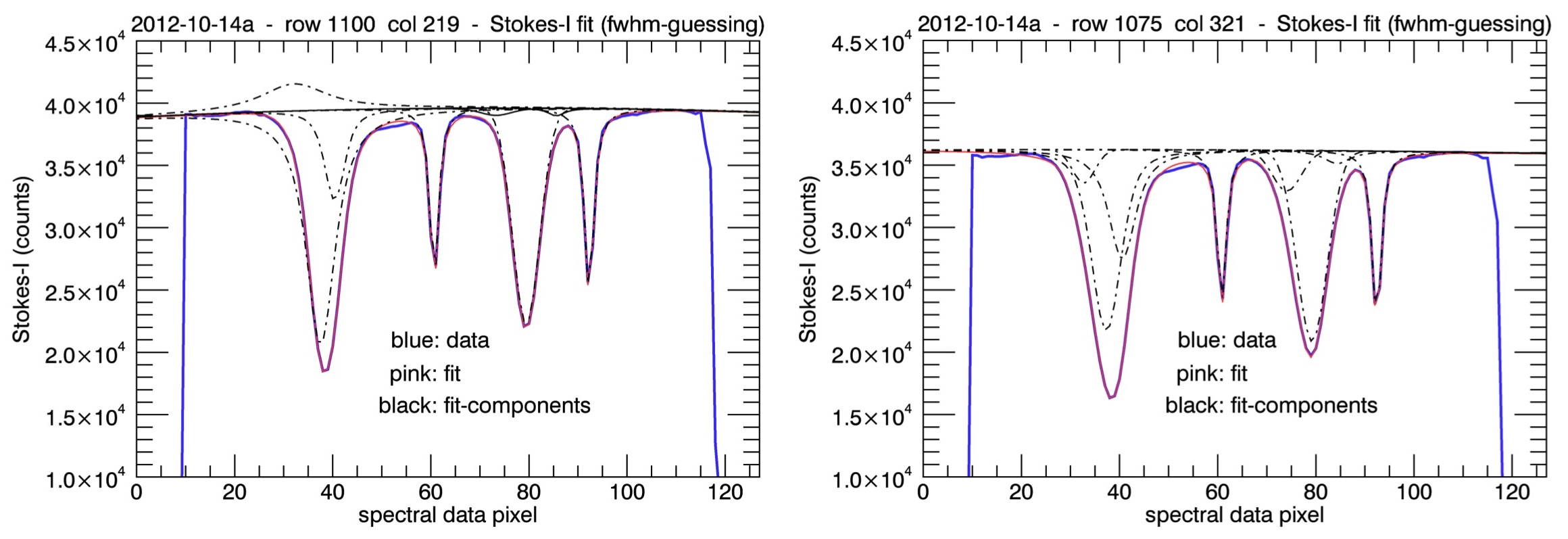}
	\caption[Unsuccessful Stokes-I fits]{Both of these Stokes-I fits belong to pixels whose reported B-value is a clear outlier (too high) from where it should be (the one in the second panel being a less-extreme outlier than the one in the first)  Note, however, \textcolor{red}{that for PROVER2B=1.2 and later} that {\em all} lines fits are now {\em forbidden} to be in emission.}
     	\label{FIG_wierdlinefits}
    	\end{center}
    	\end{figure}
	for a couple of examples.
	\item solis\_vsm\_ZEEMANFIT\_OUTFIT\_6302V() uses a fairly random threshold value to {\bf decide if the image-pixel in question is off-limb} and therefore not to fit the profile for a dispersion calculation (when running /DISPERCALC).  Since the standard is now to supply the corresponding Level-2 file for reference, the limb-darkened fit value for the solar radius could be supplied as a better measure.
	\item solis\_vsm\_ZEEMANFIT\_MULTVOIGT() computes a {\bf Lorentzian-only or Gaussian-only profile if the corresponding opposite width-parameter ($\sigma$ or $\gamma$)} is set to zero.  However, the limits suggest that, for one of these, the appropriate constraint is in approaching infinity (1/infinity = 0).  Can this be revised into a more useful/appropriate constraint or parameter definition in order to keep the fit from jumping between profile definitions if MPFITFUN() decides to sample $\sigma$ or $\gamma = 0$?
	\item The code still {\bf does not run very fast}.  Should MAXITER in MPFITFUN() be set to something smaller than the default 200?  Should the code be parallelized, taking advantage of the fact that all of the heavy lifting is done on a per-output-pixel--in-isolation basis?
	\item \_STOKESVPICK() is not used in general, but could still be necessary if someone wanted to try the Zeemanfit code out on some area scans.  However, {\bf \_STOKESVPICK() relies on hard-coded counts thresholds} (different for 6302v versus 6302l spectra) as one of the criteria for selecting high-field-strength Stokes-V spectra.  It would probably be advantageous to replace these hard-coded thresholds with some comparison ratio between the peak Stokes-V counts versus the maximum Stokes-I "continuum" counts for a given pixel-set of spectra.
	\end{enumerate}
\end{list}

\end{document}